\documentclass[%
 reprint,
superscriptaddress,
 amsmath,amssymb,
 aps,
]{revtex4-2}

\usepackage{graphicx}% Include figure files
\usepackage{bm}% bold math
\usepackage{subfig}
\usepackage{mathrsfs}
\usepackage{float}
\usepackage{url}
\graphicspath{{images}}

\allowdisplaybreaks

\begin{document}

\preprint{APS/123-QED}

\title{Twist-3 Generalized Parton Distribution for the Proton from Basis Light-Front Quantization}% Force line breaks with \\

\author{Ziqi Zhang}
\email{zhangziqi@impcas.ac.cn}
\affiliation{Institute of Modern Physics, Chinese Academy of Sciences, Lanzhou 730000, China}
\affiliation{School of Nuclear Science and Technology, University of Chinese Academy of Sciences, Beijing 100049, China}
\affiliation{CAS Key Laboratory of High Precision Nuclear Spectroscopy, Institute of Modern Physics, Chinese Academy of Sciences, Lanzhou 730000, China}

\author{Zhi Hu}
\email{huzhi0826@gmail.com}
\affiliation{Institute of Modern Physics, Chinese Academy of Sciences, Lanzhou 730000, China}
\affiliation{School of Nuclear Science and Technology, University of Chinese Academy of Sciences, Beijing 100049, China}
\affiliation{CAS Key Laboratory of High Precision Nuclear Spectroscopy, Institute of Modern Physics, Chinese Academy of Sciences, Lanzhou 730000, China}

\author{Siqi Xu}
\email{xsq234@impcas.ac.cn}
\affiliation{Institute of Modern Physics, Chinese Academy of Sciences, Lanzhou 730000, China}
\affiliation{School of Nuclear Science and Technology, University of Chinese Academy of Sciences, Beijing 100049, China}
\affiliation{CAS Key Laboratory of High Precision Nuclear Spectroscopy, Institute of Modern Physics, Chinese Academy of Sciences, Lanzhou 730000, China}

\author{Chandan Mondal}
\email{mondal@impcas.ac.cn}
\affiliation{Institute of Modern Physics, Chinese Academy of Sciences, Lanzhou 730000, China}
\affiliation{School of Nuclear Science and Technology, University of Chinese Academy of Sciences, Beijing 100049, China}
\affiliation{CAS Key Laboratory of High Precision Nuclear Spectroscopy, Institute of Modern Physics, Chinese Academy of Sciences, Lanzhou 730000, China}

\author{Xingbo Zhao}
\email{xbzhao@impcas.ac.cn}
\affiliation{Institute of Modern Physics, Chinese Academy of Sciences, Lanzhou 730000, China}
\affiliation{School of Nuclear Science and Technology, University of Chinese Academy of Sciences, Beijing 100049, China}
\affiliation{CAS Key Laboratory of High Precision Nuclear Spectroscopy, Institute of Modern Physics, Chinese Academy of Sciences, Lanzhou 730000, China}

\author{James P. Vary}
\email{jvary@iastate.edu}
\affiliation{Department of Physics and Astronomy, Iowa State University, Ames, IA 50011, U.S.A.}

\collaboration{BLFQ Collaboration}

\date{\today}% It is always \today, today,

\begin{abstract}
We investigate the twist-3 generalized parton distributions (GPDs) for the valence quarks of the proton within the basis light-front quantization (BLFQ) framework. We first solve for the mass spectra and light-front waved functions (LFWFs) in the leading Fock sector using an effective Hamiltonian. Using the LFWFs we then calculate the twist-3 GPDs via the overlap representation. By taking the forward limit, we also get the twist-3 parton distribution functions (PDFs), and discuss their properties. Our prediction for the twist-3 scalar PDF agrees well with the CLAS experimental extractions. 
\end{abstract}

\maketitle

%\tableofcontents

\section{\label{sec:level1}Introduction}
The structure of the proton constitutes one of the most fundamental problems in hadronic physics today \cite{boffi2007generalized,aidala2013spin,filippone2001spin,kuhn2009spin,deur2019spin,ji1997gauge}. It is well known that the deeply virtual Compton scattering (DVCS) experiment provides one way to explore the inner structure of the nucleon, and its scattering amplitude can be expressed in terms of the generalized parton distributions (GPDs) \cite{ji1997gauge,mueller1994wave}. There are many measurements of DVCS from Jefferson Lab (JLab) CLAS \cite{girod2008measurement,pisano2015single,jo2015cross,hattawy2019exploring}, JLab HALL A \cite{camacho2006scaling,mazouz2007deeply}, HERA H1 \cite{adloff2001measurement,aaron2008measurement,aaron2009deeply}, HERA ZEUS \cite{adloff2001measurement,zeus2009measurement} and HERA HERMES \cite{airapetian2012beam,airapetian2012beam2}, and more are 
anticipated in future EIC \cite{khalek2022science} and EicC \cite{anderle2021electron}.

The GPDs are functions of the longitudinal momentum fraction carried by the struck parton 
($x$) \cite{DIEHL200341,BELITSKY20051,ji1997gauge,Burkardt:2002hr,polyakov2018forces}, the longitudinal momentum transfer, known as skewness ($\xi$), and the square of total momentum transfer to the hadrons ($-t$), providing a three-dimensional picture of the hadron structure. The information encoded in the GPDs is richer than the ordinary parton distribution functions (PDFs) since PDFs only contain one-dimensional information. In the forward limit, the GPDs reduce to the ordinary PDFs. The GPDs can also be connected with the form factors (FFs), the orbital angular momentum (OAM), the charge distributions, etc.,  \cite{ji1997gauge,JAFFE1990509,PhysRevLett.99.112001,PhysRevLett.100.032004}. 

Most of the GPD-related studies concentrate on the leading twist \cite{boffi2007generalized,Kumericki:2009uq,Burkardt:2002ks}. The GPDs at sub-leading twist are suppressed and less known, but in fact these are not negligible. There are several reasons for the importance of the twist-3 GPDs. First, there are relations between the quark orbital angular momentum inside the nucleon and twist-3 GPDs \cite{ji1997gauge,hatta2012twist,JAFFE1990509}. Second, some studies show that from twist-3 GPDs, we can obtain information about the average transverse color Lorentz force acting on quarks \cite{PhysRevD.88.114502,aslan2019transverse}. Third, the twist-3 DVCS amplitude can be expressed in terms of the twist-3 GPDs through the Compton form factors \cite{guo2022twist}.

Basis light-front quantization (BLFQ) is a recently developed nonperturbative method \cite{PhysRevC.81.035205,PhysRevLett.106.061603,PhysRevD.88.065014,PhysRevD.91.105009,PhysRevD.96.016022,PhysRevLett.122.172001,LAN2022136890,PhysRevD.102.016008,PhysRevD.104.094036}, designed for solving relativistic bound state problems in quantum field theory (QFT) and obtaining observables from the eigenvector of the Hamiltonian. It has been successfully applied to compute many twist-2 observables \cite{Xu:2021wwj,Xu:2022abw,Peng:2022lte,Lan:2019vui,Lan:2019rba,Mondal:2019yph,Hu:2020arv,Mondal:2021czk,Nair:2022evk,Liu:2022fvl,Hu:2022ctr}.

In this work, we calculate the twist-3 GPDs of the proton using the light-front wave functions (LFWFs) obtained by diagonalizing an effective Hamiltonian of the proton within the BLFQ framework. The basis truncations and the Fock sector truncations are adopted to enable practical numerical calculations. We also present the twist-3 PDFs obtained by taking the forward limit of the twist-3 GPDs.

The organization of this work is as follows. We briefly introduce the BLFQ framework in Sec. \ref{sec2}. Then we give a simple introduction to the GPDs in Sec. \ref{sec3}, where we also show the overlap representation of the twist-3 GPDs. In Sec. \ref{sec4}, we present all the numerical results including the twist-3 GPDs and PDFs, and their properties including sum rules. At the end, we summarize this work and elaborate on future prospects in Sec. \ref{sec5}.

\section{\label{sec:level1}Basis Light-front Quantization} \label{sec2}
In this section, we will give a brief introduction to BLFQ. As a non-perturbative approach, BLFQ is based on the Hamiltonian formalism and exhibits the advantages of light-front dynamics. It has been successfully applied to many Quantum Electrodynamics (QED) and Quantum Chromodynamics (QCD) systems \cite{Li:2015zda,Zhao:2014xaa,Zhao:2022xgv,Hu:2022ctr,Kuang:2022vdy,Mondal:2021czk,Hu:2020arv,Fu:2020ksc,Hu:2019hjx,Lan:2019img,Mondal:2019jdg,Lan:2019rba,Peng:2022lte,Xu:2021wwj}.

The main idea of BLFQ is to obtain the mass spectrum and bound state wave functions simultaneously by solving the eigenvalue problem 
\begin{equation}
    H |\psi \rangle = H_\psi |\psi \rangle ,
\end{equation}
where $H$ is the light-front Hamiltonian of the system, $H_\psi$ is the eigenvalue that represents the light-front energy of the system, and $|\psi \rangle$ is the eigenvector of the system that encodes the structural information of the bound state. The notation for an arbitrary four-vector in the light-cone coordinate that we adopt is $a=(a^+,a^1,a^2,a^-)$. The $+$ and $-$ components are defined by $a^{\pm} = a^0 \pm a^3$, and the transverse directions are $\vec{a}^{\perp}=(a^1,a^2)$. 

The invariant mass is related to the light-front energy according to
\begin{equation}
    M^2 = P^+ P^- - (\vec{P}^\perp)^2 ,
\end{equation}
where $P^+$ represents the longitudinal momentum, $P^-$ represents the energy and $P^\perp$ represents the transverse momentum of the system. This leads to with $H=P^2$
\begin{equation}
    P^2 |\psi \rangle = M^2 |\psi \rangle .
\end{equation}

BLFQ uses a Fock-space expansion for hadronic states. At fixed light-front time ($x^{+} = x^{0} + x^{3} = 0$), one can expand the proton state as
\begin{equation}
    |\psi_{\mathrm{proton}} \rangle = a|qqq \rangle + b|qqqg \rangle + c|qqqq\bar{q} \rangle + \cdots ,
\end{equation}
where $\cdots$ represents all other possible parton combinations that can be found inside the proton. This work takes only the leading Fock sector into consideration. 

Within each Fock space, the proton eigensolution $|\psi (P,\Lambda) \rangle$ can be expanded in terms of $N$-parton states $|p^{+}_{i},\vec{p}^{\perp}_{i},\lambda_{i} \rangle$ as:
\begin{align}
    |\psi (P,\Lambda) \rangle =& \sum\limits_{N,\lambda_{i}} \int \prod\limits_{i=1}^{N} \frac{dx_{i} d^{2} \vec{k}^{\perp}_{i}}{\sqrt{x_{i}} 16\pi^{3}} 16\pi^{3} \delta \bigg(1 - \sum\limits_{i=1}^{N} x_{i} \bigg) \notag \\
    \times& \quad \delta^{(2)} \bigg(\sum\limits_{i=1}^{N} \vec{k}^{\perp}_{i} \bigg) \psi_{N} (x_i,\vec{k}^{\perp}_{i},\lambda_{i}) |p^{+}_{i},\vec{p}^{\perp}_{i},\lambda_{i} \rangle , \label{eq:expasion_of_psi_proton}
\end{align}
where $x_{i} = p^{+}_{i}/P^{+}$ is the longitudinal momentum fraction, $\vec{k}^{\perp}_{i} = \vec{p}^{\perp}_{i} - x_{i}\vec{P}^{\perp}$ is the transverse relative momentum, $\lambda_{i}$ is the light-cone helicity for the $i$-th parton, and $\psi_{N} (x_i,\vec{k}^{\perp}_{i},\lambda_{i})$ is the LFWF of the corresponding $N$-parton state. Those $N$-parton states are normalized as 
\begin{align}
    \langle N; \boldsymbol{p}'_{i}, \lambda'_{i} \mid N; \boldsymbol{p}_{i}, \lambda_{i} \rangle &= \prod\limits^{N}_{i=1} 16\pi^{3} p^{+}_{i} \delta (p'^{+}_{i}-p^{+}_{i}) \notag \\ 
    & \times \delta^2 (\vec{p'}^{\perp}_{i}-\vec{p}^{\perp}_{i}) \delta_{\lambda'_{i} \lambda_{i}}.
\end{align}

In this work, the light-front effective Hamiltonian designed for the proton in leading Fock sector is \cite{Xu:2021wwj}
\begin{equation}
    H_{\mathrm{eff}} = \sum_{i} \frac{m_{i}^{2} + \vec{p}_{i\perp}^{2}}{x_{i}} + \frac{1}{2} \sum_{i,j} V_{i,j}^{\mathrm{conf}} + \frac{1}{2} \sum_{i,j} V_{i,j}^{\mathrm{OGE}} ,
\end{equation}
where $m_{i}$ is the mass of $i$-th constituent, the subscript $i,j$ are the indexes for particles in the Fock sector, $V_{i,j}^{\mathrm{conf}}$ is the confining potential, and $V_{i,j}^{\mathrm{OGE}}$ is the one-gluon exchange (OGE) interaction. The specific form of the confining potential is \cite{Li:2015zda,Xu:2021wwj}
\begin{equation}
    V_{i,j}^{\mathrm{conf}} = \kappa^{4} \vec{r}^{2}_{ij\perp} + \frac{\kappa^{4}}{(m_{i} + m_{j})^{2}} \partial_{x_{i}} (x_{i} x_{j} \partial_{x_{j}}) ,
\end{equation}
where $\kappa$ defines the strength of the confining potential, $\vec{r}_{ij\perp} = \sqrt{x_{i} x_{j}} (\vec{r}_{i\perp} - \vec{r}_{j\perp})$ is the relative coordinate. The schematic form of the OGE potential \cite{Li:2015zda,Wiecki:2014ola} (see Appendix \ref{oge})
\begin{equation}
    V_{i,j}^{\mathrm{OGE}} = \frac{4\pi C_{F} \alpha_{s}}{Q^{2}_{ij}} \bar{u}_{s'_{i}} (p'_{i}) \gamma^{\mu} u_{s_{i}} (p_{i}) \bar{u}_{s'_{j}} (p'_{j}) \gamma_{\mu} u_{s_{j}} (p_{j}),
\end{equation}
where $C_{F} = -2/3$ is the color factor, $\alpha_{s}$ is the coupling constant, $Q^{2}_{ij} = -q^{2}$ is the average momentum transfer squared carried by the exchanged gluon, and $u_{s}(p)$ represents the spinor which is the solution of free Dirac equation with momentum $p$ and spin $s$. The OGE interaction plays an important role in the dynamical spin structure of LFWFs \cite{PhysRevD.104.094036}, which allows us to calculate spin dependent observables.

For the longitudinal component of the basis state, we choose a plane-wave state confined in a longitudinal box as
\begin{equation}
    \Psi_{k}(x^-) = \frac{1}{2L} e^{i \frac{\pi}{L} k \cdot x^-} ,
\end{equation}
with anti-periodic boundary condition for fermions. $2L$ is the length of the confining longitudinal box, and $k$ is the quantum number that represents the longitudinal degree of freedom. The longitudinal momentum is given by $p^+ = 2\pi k / 2 L$, with $k=\{1/2,3/2,5/2,\cdots\}$. Two-dimensional harmonic oscillator (2D-HO) states are adopted for the transverse components as
\begin{equation}
    \phi^m_n(\rho, \varphi) = \frac{1}{b} \sqrt{\frac{4\pi n!}{(|m|+n)!}} e^{im\varphi} \rho^{|m|} e^{-\rho^2/2} L^{|m|}_n(\rho^2) ,
\end{equation}
where $n$, $m$ are the radial and angular quantum numbers, respectively, $\rho \equiv |p|/b$ is a dimensionless argument, $b$ is the basis scale parameter which has mass dimension, $L^{|m|}_n$ is the associated Laguerre polynomial, and $\varphi = \arg(\vec{p}_{\perp})$. The last degree of freedom is the light-cone helicity state in the spin space represented by $\lambda$. Then we have a complete set of single-particle quantum numbers that represent a Fock-particle state, $\{k,n,m,\lambda\}$. The basis states of a Fock sector with multiple particles are expressed as the direct product of each single-particle basis state in the Fock sector. These single-particle states are orthonormal.

Within the BLFQ framework, we introduce a Fock-sector truncation and two cutoffs for practical calculations. Only the first Fock sector $|qqq \rangle$ is taken into consideration in this work. The two cutoffs are represented by $N_{\mathrm{max}}$ and $K$. $N_{\mathrm{max}}$ is the cutoff in the total energy of the 2D-HO basis states in the transverse direction, given by $\sum_i (2n_i + |m_i| + 1) \leq N_{\mathrm{max}}$, and $K = \Sigma_{i} k_{i}$ represents the cutoff in the longitudinal direction.

This work uses single-particle states to construct the BLFQ basis. It has an advantage for retaining the correct fermion statistics for quarks. However, the many-particle basis therefore incorporates the transverse center-of-mass (CM) motion which is entangled with intrinsic motion. It is then necessary to add a constraint term into the effective Hamiltonian
\begin{equation}
    H' = \lambda_{L} (H_{CM} - 2b^{2} I) ,
\end{equation}
to enforce factorization of LFWFs into a product of internal motion and CM motion components.  By removing the transverse CM motion component, one obtains a boost-invariant LFWF. 

The CM motion is governed by 
\begin{equation}
    H_{CM} = \bigg( \sum\limits_{i} \vec{p}^{\perp}_{i} \bigg)^{2} + b^{4} \bigg( \sum\limits_{i} x_{i} \vec{r}^{\perp}_{i} \bigg)^{2} .
\end{equation}
Here $\lambda_{L}$ is the Lagrange multiplier, $2b^{2}$ is the zero-point energy and $I$ is a unity operator. By setting $\lambda_{L}$ sufficiently large, it is possible to shift the excited states of CM motion to higher energy and ensure that low-lying states are all in the ground state of CM motion.

The LFWF of the proton state is obtained in terms of BLFQ LFWF as 
\begin{align}
    \psi_{\{x_{i},p_{i}^{\perp},\lambda_{i}\}}^{\Lambda} &= \langle \{p_{i}^{+},p_{i}^{\perp},\lambda_{i}\} | \psi(P,\Lambda) \rangle \notag \\
    &= \sum\limits_{\{n_i,m_i\}} \psi_{\{x_{i},n_{i},m_{i},\lambda_{i}\}}^{\Lambda} \prod \phi_{n_{i}}^{m_{i}}(\vec{p}_{i}),
\end{align}
where the completeness $\sum |\{x,n,m,\lambda\} \rangle \langle \{x,n,m,\lambda\}|=1$ is used, and the $\psi_{\{x_{i},n_{i},m_{i},\lambda_{i}\}}^{\Lambda}=\langle \{x_{i},n_{i},m_{i},\lambda_{i}\} | \psi(P,\Lambda) \rangle$ are the wave functions in BLFQ which can be obtained from diagonalizing the light-front Hamiltonian.

All the calculations below are performed with the following set of parameters: the basis truncation $N_{\mathrm{max}} = 10$ and $K = 16.5$, the model parameters $m_{q/\mathrm{KE}} = 0.3 \ \mathrm{GeV}$, $m_{q/\mathrm{OGE}} = 0.2 \ \mathrm{GeV}$, coupling constant $\alpha_{s} = 0.55$, confining potential $\kappa=0.34\ \mathrm{GeV}$ and 2D-HO scale parameter $b = 0.6 \ \mathrm{GeV}$. The proton LFWFs resulting with this parameter set have been successfully applied to compute a wide class of different and related proton obsevables, e.g., the electromagnetic and axial form factors, radii, leading twist PDFs, GPDs, helicity asymmetries, TMDs, etc, with remarkable overall success \cite{Xu:2021wwj,Xu:2022abw,Peng:2022lte,Lan:2019vui,Lan:2019rba,Mondal:2019yph,Hu:2020arv,Mondal:2021czk,Nair:2022evk,Liu:2022fvl,Hu:2022ctr,Kaur:2023lun}.

\section{\label{sec:level1}Generalized Parton Distribution} \label{sec3}
In this section, we will present the definition of twist-3 GPDs, and their overlap representations using LFWFs. The GPDs are functions of three variables, $x$ representing the longitudinal fraction, $\xi$ for the skewness, and $-t$ signifying the momentum transfer squared. The GPDs are defined as off-forward matrix elements of a bilocal operator as
\begin{widetext}
\begin{equation}
    F^{[\Gamma]}_{\Lambda'\Lambda} (x,\xi,t) = \frac{1}{2} \int \frac{dy^{-}}{4\pi} e^{ixP\cdot y} \bigg\langle P',\Lambda' \bigg| \bar{\Psi}\left(-\frac{y}{2}\right) \mathscr{W}\left(-\frac{y}{2},\frac{y}{2}\right) \Gamma \Psi\left(\frac{y}{2}\right) \bigg| P,\Lambda \bigg\rangle \bigg|_{y^{+}=0,\vec{y}^{\perp}=0} , \label{gpddef}
\end{equation}
\end{widetext}
where $P$, $P'$ represent the momenta of the initial and final proton respectively, $\Lambda$, $\Lambda'$ represent the light-front helicities of the initial and final proton respectively, and $\mathscr{W}(y,x) \equiv P \mathrm{exp}(ig\int^y_x dz^\mu A_\mu(z))$ is the gauge link that ensures that the bilocal operator remains gauge invariant. Since we are working in the light-cone gauge (where $A^{+}=0$), the gauge link is then unity. $\Gamma$ is one of the sixteen Dirac gamma matrices. We choose a symmetric frame throughout this work:
\begin{align}
    P =& \bigg( (1+\xi)\bar{P}^+,-\frac{\vec{\Delta}^{\perp}}{2},\frac{M^2 + (\vec{\Delta}^\perp)^2/4}{(1+\xi)\bar{P}^+} \bigg) , \\
    P' =& \bigg( (1-\xi)\bar{P}^+,\frac{\vec{\Delta}^{\perp}}{2},\frac{M^2 + (\vec{\Delta}^\perp)^2/4}{(1-\xi)\bar{P}^+} \bigg) , \\
    \Delta =& P' - P = \bigg( -2\xi \bar{P}^+ , \vec{\Delta}^\perp , \frac{t - (\vec{\Delta}^\perp)^2}{2\xi \bar{P}^+} \bigg) ,
\end{align}
where $\bar{P}=(P+P')/2$ is the average momentum, $\Delta$ is the momentum transfer, $\xi = -\Delta^{+}/2\bar{P}^{+}$ is the skewness, $M$ is the proton mass, and $t\equiv -\Delta^2$ is the momentum transfer squared. Usually there are two $\xi$-dependent regions in the DVCS process for quarks, one is $-\xi < x < \xi$, called the Efremov-Radyushkin-Brodsky-Lepage (ERBL) region, the other is $\xi < |x| < 1$, called the Dokshitzer-Gribov-Lipatov-Altarelli-Parisi (DGLAP) region. This work focuses on the zero-skewness limit, i.e. $\xi=0$, so only the DGLAP region applies. The DGLAP region describes a quark scattered off the proton, absorbing the virtual photon and immediately radiating a real photon, then returning to form a recoiled proton. This is a $n\rightarrow n$ diagonal (parton-number conserved) process.

With the parameterization taken from Ref. \cite{meissner2009generalized}, which includes both chiral-even and chiral-odd GPDs, one finds that the sixteen sub-leading twist GPDs are defined as
\begin{widetext}
\begin{align}
    F^{[\gamma^j]}_{\Lambda'\Lambda} =& \frac{M}{2(\bar{P}^+)^2} \bar{u}(P',\Lambda') \bigg[i\sigma^{+j} H_{2T}(x,\xi,-t) + \frac{\gamma^+ \vec{\Delta}^j_\perp - \Delta^+ \gamma^j}{2M} E_{2T}(x,\xi,-t)   \notag \\
    &+ \frac{\bar{P}^+ \vec{\Delta}^j_\perp - \Delta^+ \vec{\bar{P}}^j_\perp}{M^2} \tilde{H}_{2T}(x,\xi,-t) + \frac{\gamma^+ \vec{\bar{P}}^j_\perp - \bar{P}^+ \gamma^j}{M} \tilde{E}_{2T}(x,\xi,-t) \bigg] u(P,\Lambda), \label{gammagpd} \\
    F^{[\gamma^j \gamma_5]}_{\Lambda'\Lambda} =& -\frac{i\epsilon^{ij}_\perp}{2(\bar{P}^+)^2} \bar{u}(P',\Lambda') \bigg[i\sigma^{+i} H'_{2T}(x,\xi,-t) + \frac{\gamma^+ \vec{\Delta}^i_\perp - \Delta^+ \gamma^i}{2M} E'_{2T}(x,\xi,-t)   \notag \\
    &+ \frac{\bar{P}^+ \vec{\Delta}^i_\perp - \Delta^+ \vec{\bar{P}}^i_\perp}{M^2} \tilde{H}'_{2T}(x,\xi,-t) + \frac{\gamma^+ \vec{\bar{P}}^i_\perp - \bar{P}^+ \gamma^i}{M} \tilde{E}'_{2T}(x,\xi,-t) \bigg] u(P,\Lambda), \label{gamma5gpd} \\
    F^{[1]}_{\Lambda'\Lambda} =& \frac{M}{2(\bar{P}^+)^2} \bar{u}(P',\Lambda') \bigg[\gamma^+ H_2(x,\xi,-t) + \frac{i\sigma^{+ \rho}\Delta_{\rho}}{2M} E_2(x,\xi,-t)\bigg] u(P,\Lambda), \\
    F^{[\gamma_5]}_{\Lambda'\Lambda} =& \frac{M}{2(\bar{P}^+)^2} \bar{u}(P',\Lambda') \bigg[\gamma^+ \gamma_5 \tilde{H}_2(x,\xi,-t) + \frac{\bar{P}^+ \gamma_5}{2M} \tilde{E}_2(x,\xi,-t)\bigg] u(P,\Lambda), \label{eqgamma5} \\
    F^{[i\sigma^{ij} \gamma_5]}_{\Lambda'\Lambda} =& -\frac{i\epsilon^{ij}_\perp}{2(\bar{P}^+)^2} \bar{u}(P',\Lambda') \bigg[\gamma^+ H'_2(x,\xi,-t) + \frac{i\sigma^{+ \rho}\Delta_\rho}{2M} E'_2(x,\xi,-t)\bigg] u(P,\Lambda), \\
    F^{[i\sigma^{+-} \gamma_5]}_{\Lambda'\Lambda} =& \frac{M}{2(\bar{P}^+)^2} \bar{u}(P',\Lambda') \bigg[\gamma^ + \gamma_5 \tilde{H}'_2(x,\xi,-t) + \frac{\bar{P}^+ \gamma_5}{2M} \tilde{E}'_2(x,\xi,-t)\bigg] u(P,\Lambda), \label{eqend}
\end{align}
\end{widetext}
where $\sigma^{ij}=i[\gamma^{i},\gamma^{j}]/2$, and $\epsilon^{ij}_{\perp}=\epsilon^{-+ij}$ with the antisymmetric Levi-Civita tensor $\epsilon^{-+12} = 1$. Here $i,j$ can only be transverse indice $1,2$. 

A different parameterization of GPDs with vector and axial vector is introduced in Ref. \cite{PhysRevD.98.014038}
\begin{widetext}
\begin{align}
    F^{\mu} =& \bar{u} (P') \bigg[ \bar{P}^{\mu} \frac{\gamma^{+}}{\bar{P}^{+}} H + \bar{P}^{\mu} \frac{i\sigma^{\mu\nu} \Delta_{\nu}}{2M} E + \frac{\Delta^{\mu}_{\perp}}{2M} G_{1} + \gamma^{\mu} (H + E + G_{2}) + \frac{\Delta^{\mu}_{\perp} \gamma^{+}}{\bar{P}^{+}} G_{3} + \frac{i\epsilon^{\mu\nu}_{\perp} \Delta_{\nu} \gamma^{\mu} \gamma_{5}}{\bar{P}^{+}} G_{4} \bigg] u (P) , \\
    \tilde{F}^{\mu} =& \bar{u} (P') \bigg[ \bar{P}^{\mu} \frac{\gamma^{+} \gamma_{5}}{\bar{P}^{+}} \tilde{H} + \bar{P}^{\mu} \frac{\Delta^{\mu}}{2M} \tilde{E} + \frac{\Delta^{\mu}_{\perp} \gamma_{5}}{2M} (\tilde{E} + \tilde{G}_{1}) + \gamma^{\mu} \gamma_{5} (\tilde{H} + \tilde{G}_{2}) + \frac{\Delta^{\mu}_{\perp} \gamma^{+} \gamma_{5}}{\bar{P}^{+}} \tilde{G}_{3} + \frac{i\epsilon^{\mu\nu}_{\perp} \Delta_{\nu} \gamma^{\mu}}{\bar{P}^{+}} \tilde{G}_{4} \bigg] u (P) .
\end{align}
\end{widetext}

By using relations based on the Dirac equation \cite{Belitsky2005UnravelingHS,BELITSKY2000611,PhysRevD.97.016005,PhysRevD.98.014038}, the two types of GPDs above can actually be related to each other according to the equations in Appendix \ref{app}. In addition, there is also another parameterization defined in Ref. \cite{guo2021novel},
\begin{align}
    F_{q,\gamma^{\perp}} &= \frac{\vec{\Delta}^{\perp}}{M} G_{q,1} (x,\xi,-t) + \vec{\Delta}^{\perp} n\! \! \! / G_{q,2} (x,\xi,-t) \notag \\
    &+ \frac{i\sigma^{\perp\rho} \Delta_{\rho}}{2M} G_{q,3} (x,\xi,-t) + M i\sigma^{\perp\rho} n_{\rho} G_{q,4} (x,\xi,-t), \label{par2}
\end{align}
and their relations to the GPDs in this work can also be found in Appendix \ref{app}. There are also some other parameterizations of GPDs \cite{belitsky2001twist,belitsky2002theory,PhysRevD.88.014041,PhysRevD.100.096021}, but we will not illustrate them here.

We will present the overlap representations of all zero-skewness twist-3 GPDs in the following. For convenience, the following notations shall be used,
\begin{align}
    [dx] [d^2 k] &= \prod \frac{d x_{i} d^2 \vec{k}_{i}}{16 \pi^3} 16\pi^3 \delta \left(1-\sum x_i \right) \notag \\ 
    &\times \delta^2 \left(\sum \vec{k}_i \right) \delta (x-x_1),
\end{align}
and $[\Gamma]=\bar{u} (p',\lambda') \Gamma u (p,\lambda)$ encodes struck quark helicity combinations. Also $\psi^{\Lambda}_{\lambda_1}$ signifies the LFWF $\psi^{\Lambda}_{\lambda_1,\lambda_2,\lambda_3}(x_i,p_i,\lambda_i)$, where $\Lambda$ represents the proton helicity, and $\lambda_1$ represents the struck quark helicity. The constraint of spectators $\delta^{\lambda'_2}_{\lambda_2} \delta^{\lambda'_3}_{\lambda_3}$ is implied. In the overlap expressions, the symbol $\Delta$ will refer to the 2D complex representation, $\Delta = \Delta_{1} + i\Delta_{2}$. By taking $\Gamma=\gamma^{\perp}$, one finds the following expressions
\begin{align}
    H^{\gamma^j}_{2T}(x,0,-t) &= \sum_{\lambda_{i}} \int [dx] [d^2k] [\Gamma] \frac{P^{+}}{2M} \frac{(-)^{j} 2 i}{\Delta + (-)^{j} \Delta^{\star}} \notag \\ 
    &\times \bigg( \Delta \psi^{\uparrow \star}_{\lambda'_1} \psi^{\downarrow}_{\lambda_1} + \Delta^{\star} \psi^{\downarrow \star}_{\lambda'_1} \psi^{\uparrow}_{\lambda_1} \bigg), \label{olpst} \\
    E^{\gamma^j}_{2T}(x,0,-t) &= \sum_{\lambda_{i}} \int [dx] [d^2k] [\Gamma] \notag \\
    &\times \bigg( \frac{2 P^{+}}{\Delta - (-)^{j}\Delta^{\star}} \big( \psi^{\uparrow \star}_{\lambda'_1} \psi^{\uparrow}_{\lambda_1} + \psi^{\downarrow \star}_{\lambda'_1} \psi^{\downarrow}_{\lambda_1} \big) \notag \\ 
    &+ \frac{(-)^{j} 2(i)^{j}MP^{+}}{\mathrm{Re}(\Delta) \mathrm{Im}(\Delta)} \big( \psi^{\uparrow \star}_{\lambda'_1} \psi^{\downarrow}_{\lambda_1} - (-)^{j} \psi^{\downarrow \star}_{\lambda'_1} \psi^{\uparrow}_{\lambda_1} \big) \bigg), \\
    \tilde{H}^{\gamma^j}_{2T}(x,0,-t) &= \sum_{\lambda_{i}} \int [dx] [d^2k] [\Gamma] \frac{- (-i)^{j} MP^{+}}{\mathrm{Re}(\Delta) \mathrm{Im}(\Delta)} \notag \\ 
    &\times \bigg( \psi^{\uparrow \star}_{\lambda'_1} \psi^{\downarrow}_{\lambda_1} - (-)^{j} \psi^{\downarrow \star}_{\lambda'_1} \psi^{\uparrow}_{\lambda_1} \bigg), \\
    \tilde{E}^{\gamma^j}_{2T}(x,0,-t) &= \sum_{\lambda_{i}} \int [dx] [d^2k] [\Gamma] \frac{(-)^{j} 2iP^{+}}{\Delta + (-)^{j}\Delta^{\star}} \notag \\ 
    &\times \bigg( \psi^{\uparrow \star}_{\lambda'_1} \psi^{\uparrow}_{\lambda_1} - \psi^{\downarrow \star}_{\lambda'_1} \psi^{\downarrow}_{\lambda_1} \bigg).
\end{align}
With $\Gamma=\gamma^{\perp}\gamma_{5}$, the expressions are 
\begin{align}
    H'^{\gamma^{j} \gamma_{5}}_{2T}(x,0,-t) &= \sum_{\lambda_{i}} \int [dx] [d^2k] [\Gamma] \frac{P^{+}}{2M} \frac{2}{\Delta - (-)^{j} \Delta^{\star}} \notag \label{gth2tp} \\ 
    &\times \bigg( \Delta \psi^{\uparrow \star}_{\lambda'_1} \psi^{\downarrow}_{\lambda_1} + \Delta^{\star} \psi^{\downarrow \star}_{\lambda'_1} \psi^{\uparrow}_{\lambda_1} \bigg), \\
    E'^{\gamma^{j} \gamma_{5}}_{2T}(x,0,-t) &= \sum_{\lambda_{i}} \int [dx] [d^2k] [\Gamma] \notag \\
    &\times \bigg( \frac{(-)^{j} 2i P^{+}}{\Delta + (-)^{j} \Delta^{\star}} \big( \psi^{\uparrow \star}_{\lambda'_1} \psi^{\uparrow}_{\lambda_1} + \psi^{\downarrow \star}_{\lambda'_1} \psi^{\downarrow}_{\lambda_1} \big) \notag \\ 
    &+ \frac{- 2(-i)^{j} MP^{+}}{\mathrm{Re}(\Delta) \mathrm{Im}(\Delta)} \big( \psi^{\uparrow \star}_{\lambda'_1} \psi^{\downarrow}_{\lambda_1} + (-)^{j} \psi^{\downarrow \star}_{\lambda'_1} \psi^{\uparrow}_{\lambda_1} \big) \bigg), \\
    \tilde{H}'^{\gamma^j \gamma_{5}}_{2T}(x,0,-t) &= \sum_{\lambda_{i}} \int [dx] [d^2k] [\Gamma] \frac{(-i)^{j} MP^{+}}{\mathrm{Re}(\Delta) \mathrm{Im}(\Delta)} \notag \\
    &\times \bigg( \psi^{\uparrow \star}_{\lambda'_1} \psi^{\downarrow}_{\lambda_1} + (-)^{j} \psi^{\downarrow \star}_{\lambda'_1} \psi^{\uparrow}_{\lambda_1} \bigg), \\
    \tilde{E}'^{\gamma^j \gamma_5}_{2T}(x,0,-t) &= \sum_{\lambda_{i}} \int [dx] [d^2k] [\Gamma] \frac{2P^{+}}{\Delta - (-)^{j}\Delta^{\star}} \notag \\
    &\times \bigg( \psi^{\uparrow \star}_{\lambda'_1} \psi^{\uparrow}_{\lambda_1} - \psi^{\downarrow \star}_{\lambda'_1} \psi^{\downarrow}_{\lambda_1} \bigg). \label{olf8ed}
\end{align}
For $\Gamma=1$, they are
\begin{align}
    H_{2}(x,0,-t) &= \sum_{\lambda_{i}} \int [dx] [d^2p] [\Gamma] \frac{P^{+}}{M} \psi^{\uparrow \star}_{\lambda'_1} \psi^{\uparrow}_{\lambda_1} , \label{exh2} \\
    E_{2}(x,0,-t) &= \sum_{\lambda_{i}} \int [dx] [d^2p] [\Gamma] \frac{-2P^{+}}{\Delta^{\star}} \psi^{\uparrow \star}_{\lambda'_1} \psi^{\downarrow}_{\lambda_1} .
\end{align}
For $\Gamma=\gamma_{5}$,
\begin{align}
    \tilde{H}_{2}(x,0,-t) &= \sum_{\lambda_{i}} \int [dx] [d^2p] [\Gamma] \frac{P^{+}}{M} \psi^{\uparrow \star}_{\lambda'_1} \psi^{\uparrow}_{\lambda_1} , \\
    \tilde{E}_{2}(x,0,-t) &= \sum_{\lambda_{i}} \int [dx] [d^2p] [\Gamma] \frac{-2P^{+}}{\Delta^{\star}} \psi^{\uparrow \star}_{\lambda'_1} \psi^{\downarrow}_{\lambda_1} .
\end{align}
For $\Gamma=i \sigma^{ij} \gamma_{5}$,
\begin{align}
    H'_{2}(x,0,-t) &= \sum_{\lambda_{i}} \int [dx] [d^2p] [\Gamma] \frac{i P^{+}}{M} \psi^{\uparrow \star}_{\lambda'_1} \psi^{\uparrow}_{\lambda_1} , \label{hlh2tp} \\
    E'_{2}(x,0,-t) &= \sum_{\lambda_{i}} \int [dx] [d^2p] [\Gamma] \frac{-2 i P^{+}}{\Delta^{\star}} \psi^{\uparrow \star}_{\lambda_1} \psi^{\downarrow}_{\lambda'_1} .
\end{align}
For $\Gamma=i \sigma^{+-} \gamma_5$,
\begin{align}
    \tilde{H}'_{2}(x,0,-t) &= \sum_{\lambda_{i}} \int [dx] [d^2p] [\Gamma] \frac{P^{+}}{M} \psi^{\uparrow \star}_{\lambda'_1} \psi^{\uparrow}_{\lambda_1} , \\
    \tilde{E}'_{2}(x,0,-t) &= \sum_{\lambda_{i}} \int [dx] [d^2p] [\Gamma] \frac{-2 P^{+}}{\Delta^{\star}} \psi^{\uparrow \star}_{\lambda'_1} \psi^{\downarrow}_{\lambda_1}. \label{olped}
\end{align}

\section{\label{sec:level1}Numerical Results and Discussion} \label{sec4}
In this section, we will present the numerical results of zero-skewness twist-3 GPDs calculated by the BLFQ method, and discuss their properties including sum rules, time-reversal symmetry and PDF limits.

\subsection{\label{sec:level2}Twist-3 GPDs}
\begin{figure*}[htbp]
    \centering
    \subfloat[ ]{\includegraphics[scale=0.65]{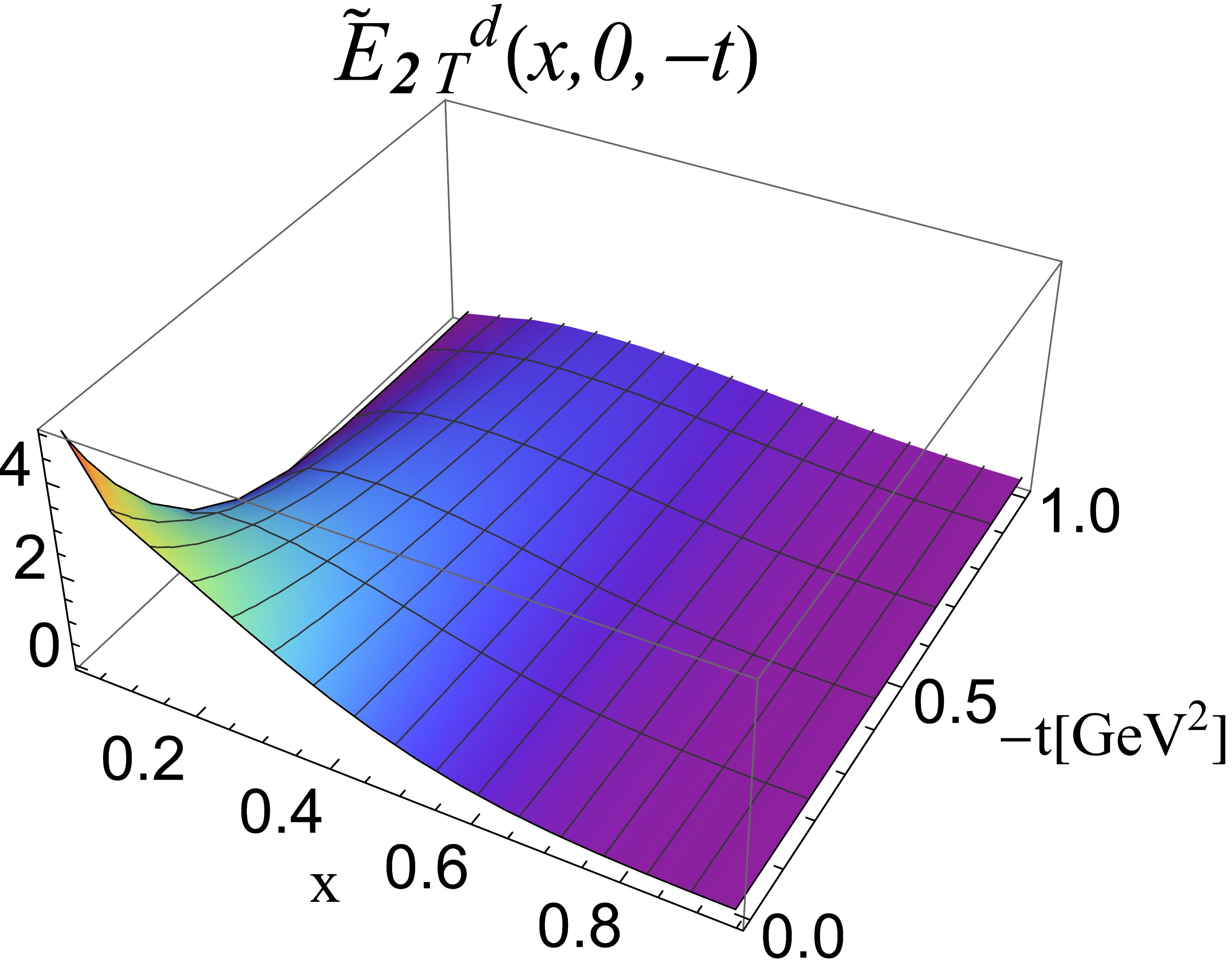}} \hspace{.3in}
    \subfloat[ ]{\includegraphics[scale=0.65]{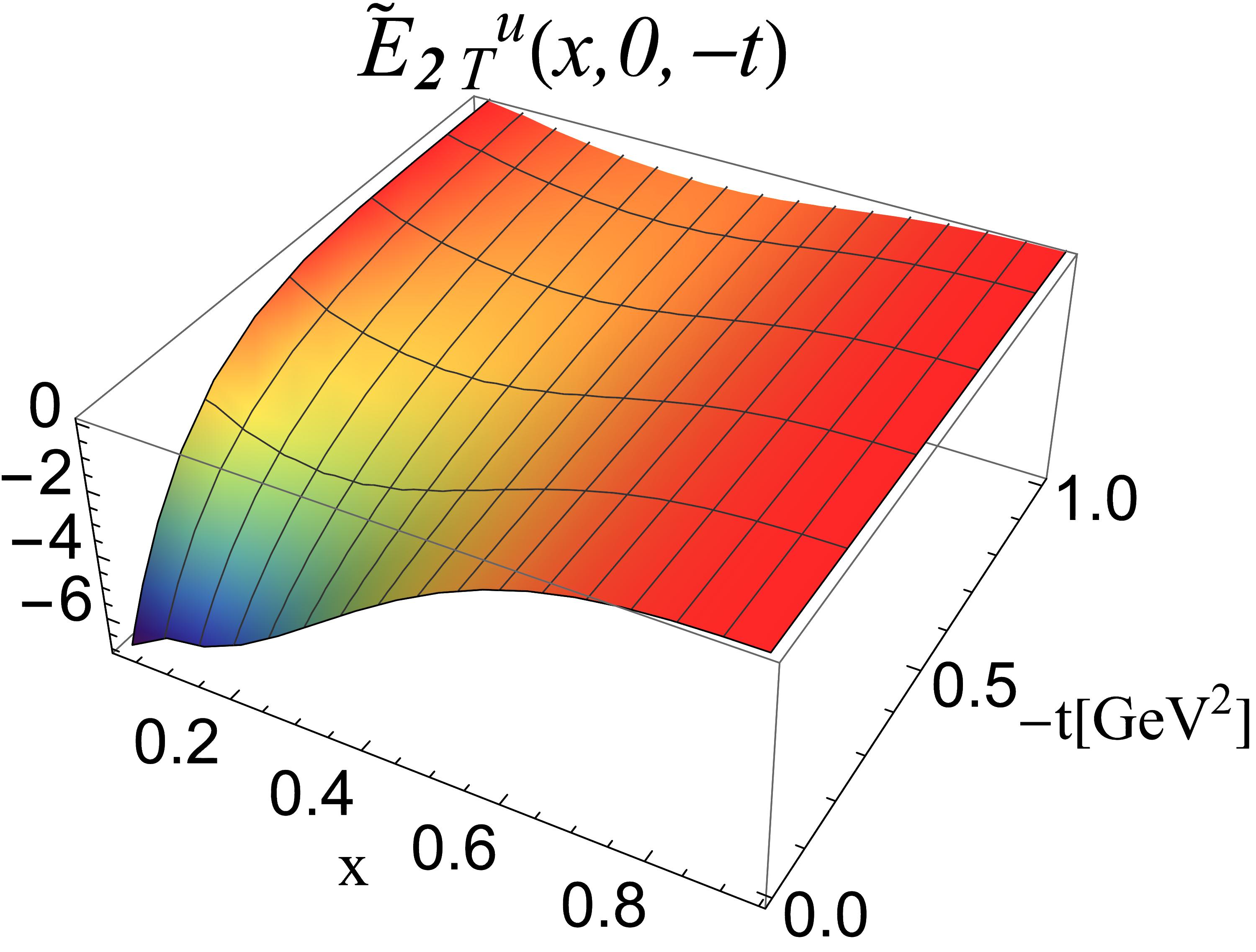}} 
    \captionsetup{justification=raggedright}
    \caption{Twist-3 quark GPDs in the proton associated with $\Gamma=\gamma^{\perp}$; (a) and (b) are for the down and up quarks on the quark level, respectively. The flavor level distributions are given by $X_{flavor}^u =2 X_{quark}^u$ and $X_{flavor}^d=X_{quark}^d$, where $X$ stands for all the GPDs. The GPDs are evaluated with $N_{\mathrm{max}}=10$ and $K=16.5$.}
    \label{plotgammamu}
\end{figure*}

\begin{figure*}[htbp]
    \centering
    \subfloat[ ]{\includegraphics[scale=0.65]{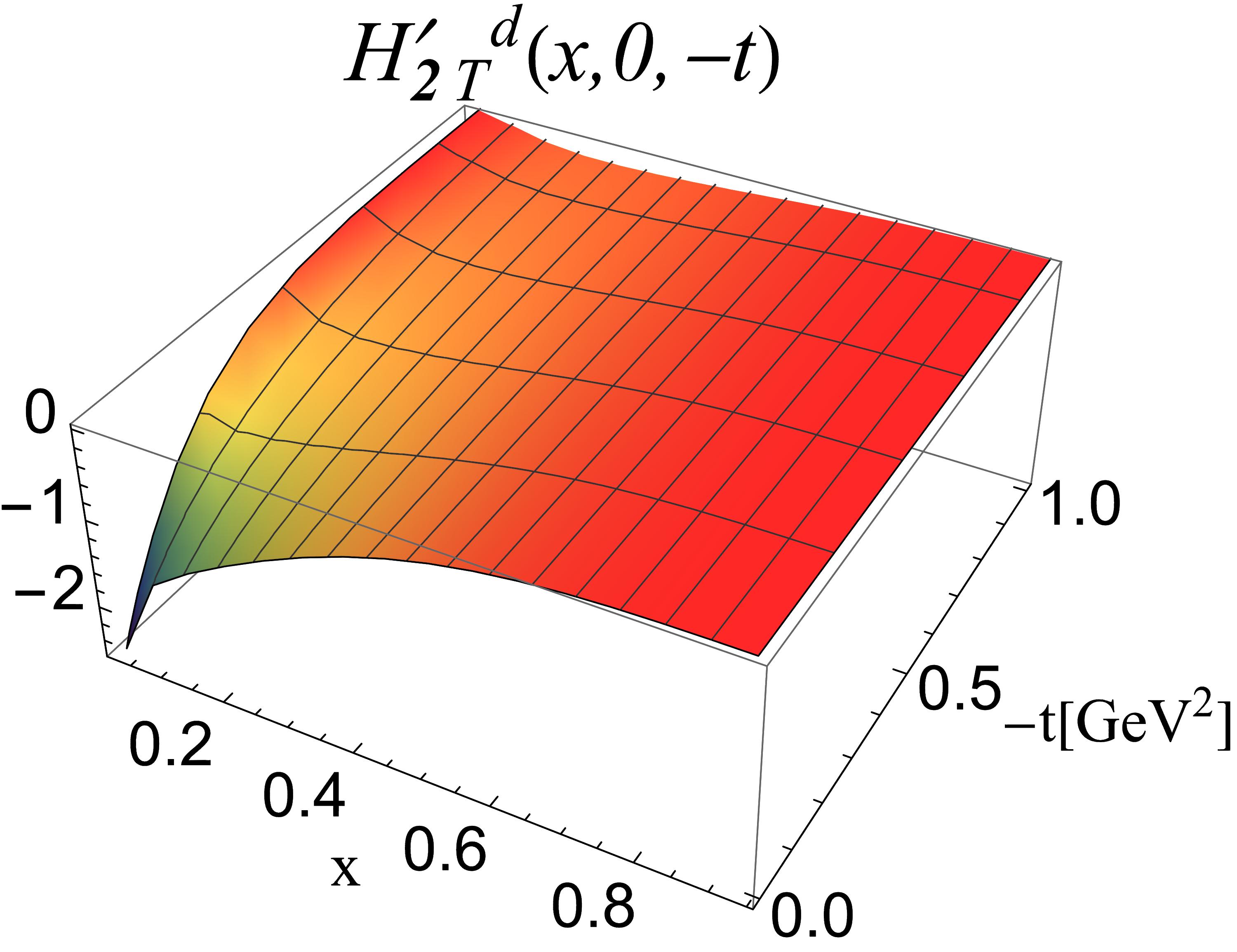}} \hspace{.3in}
    \subfloat[ ]{\includegraphics[scale=0.65]{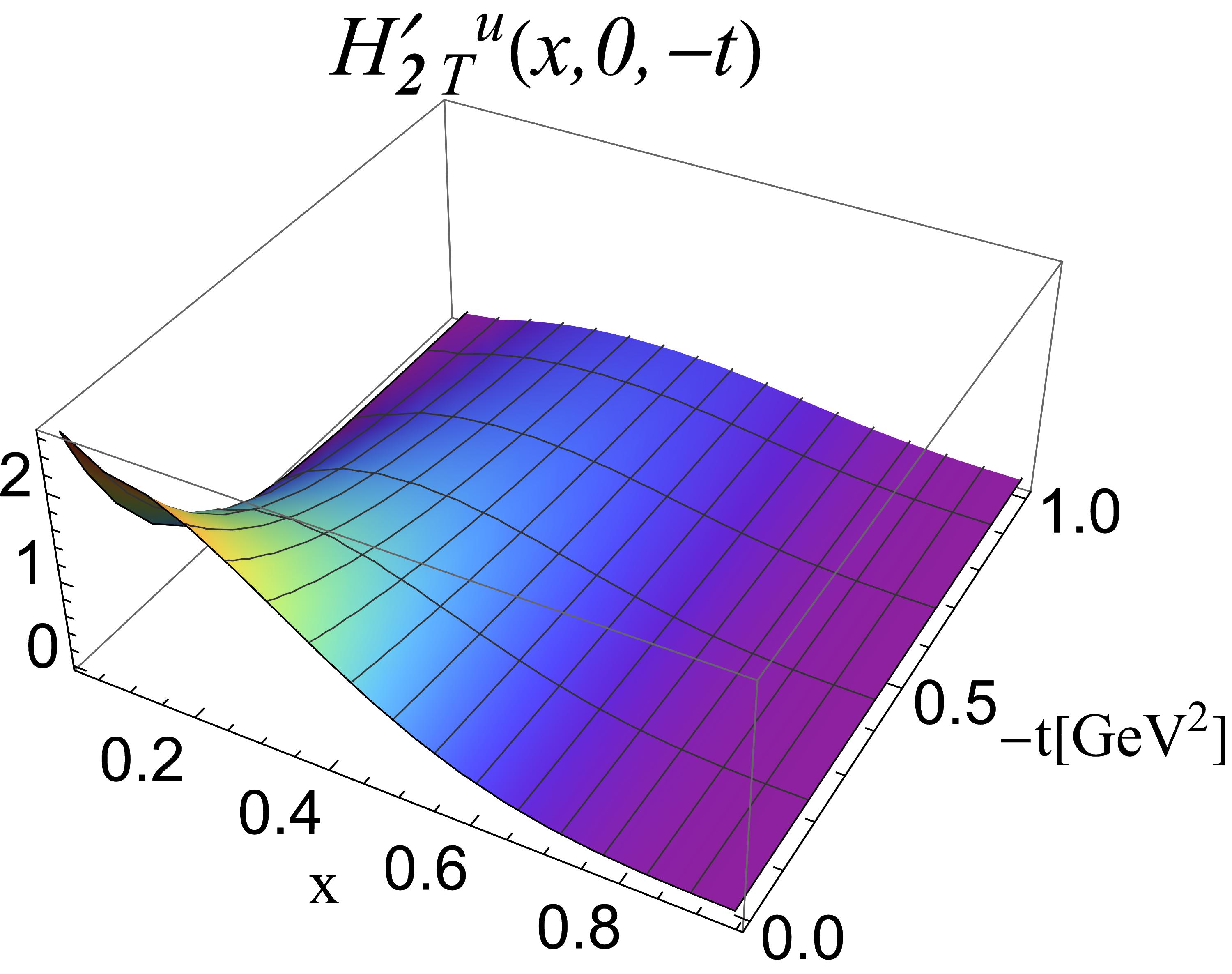}} \\
    \subfloat[ ]{\includegraphics[scale=0.65]{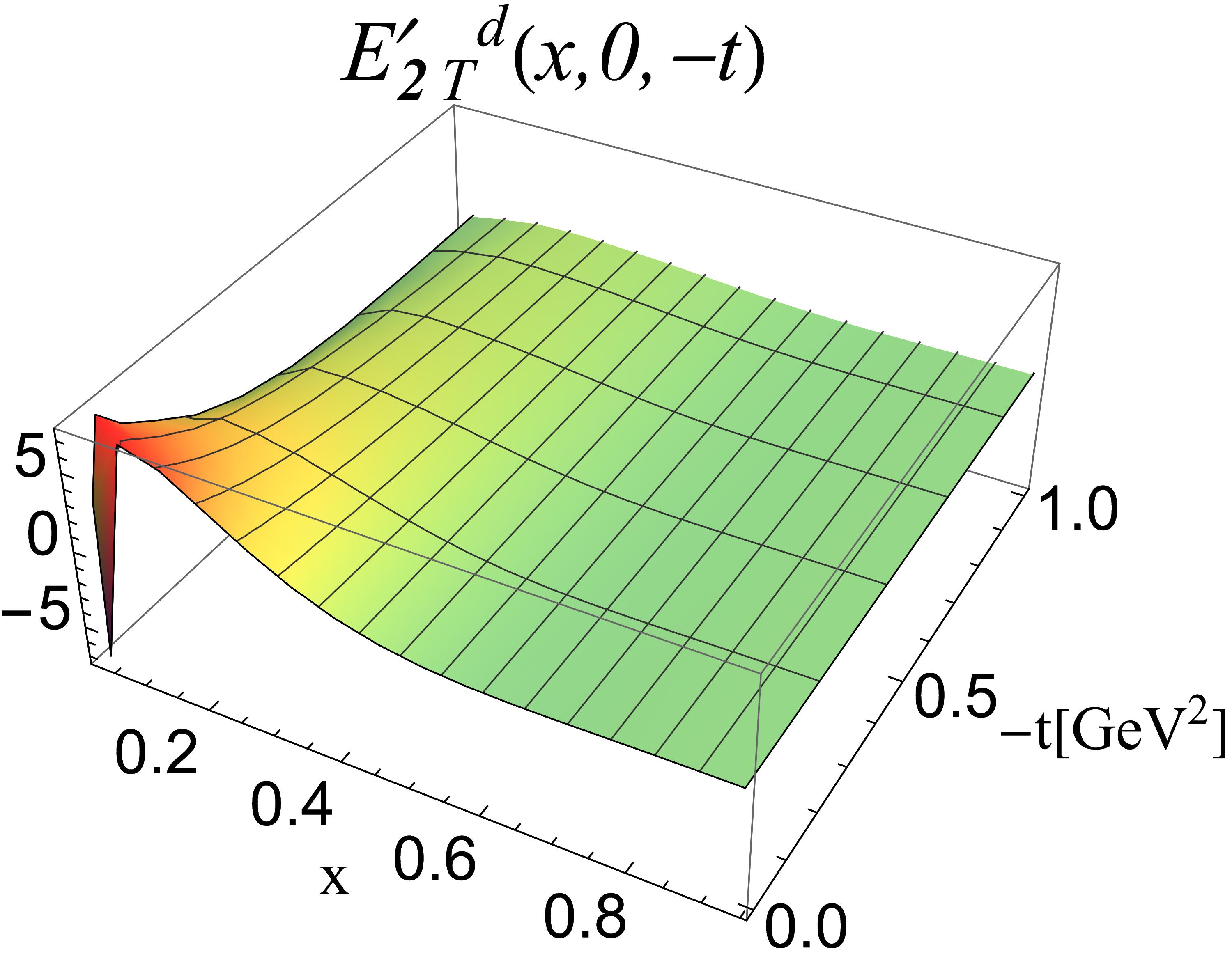}} \hspace{.3in}
    \subfloat[ ]{\includegraphics[scale=0.65]{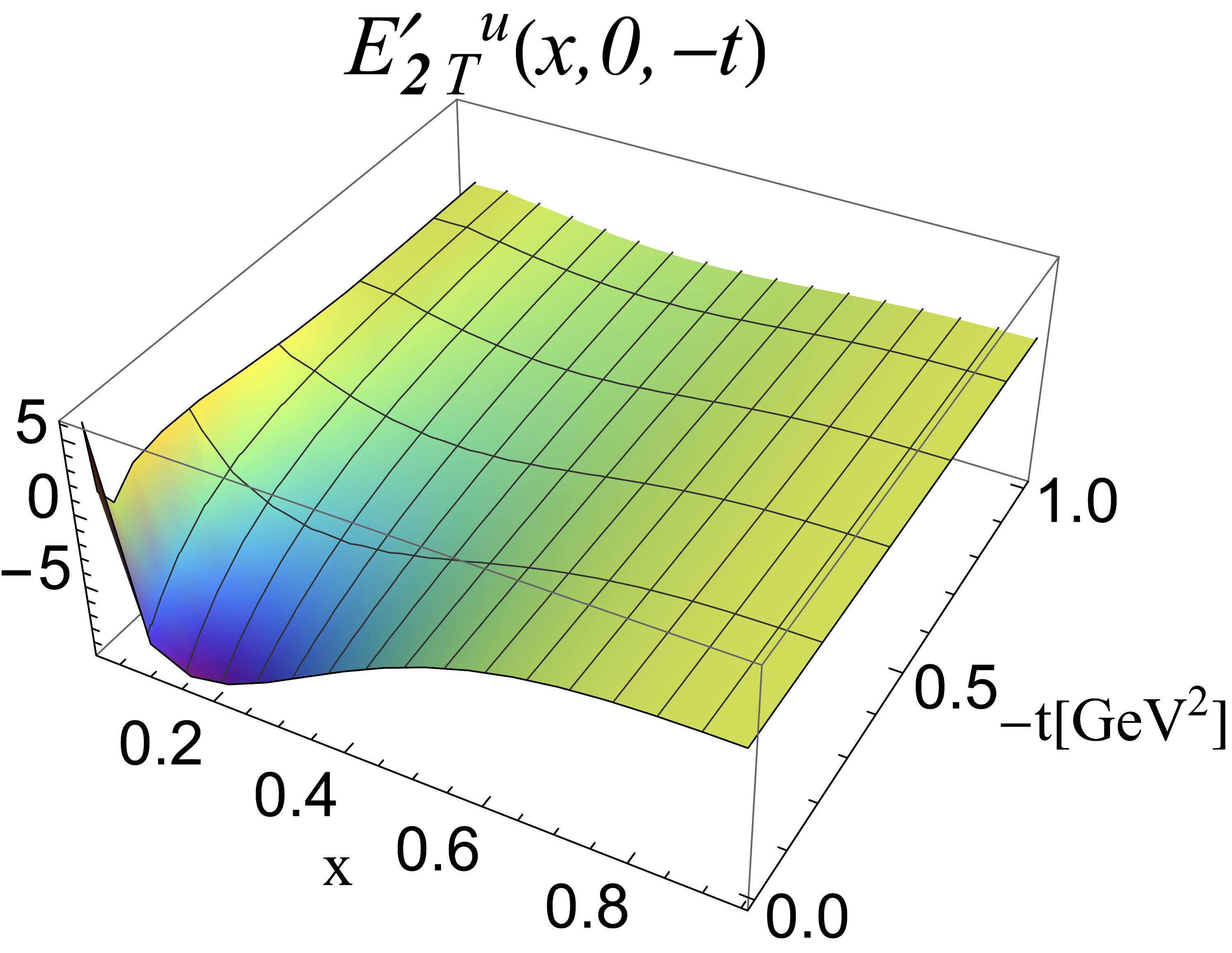}} \\
    \subfloat[ ]{\includegraphics[scale=0.65]{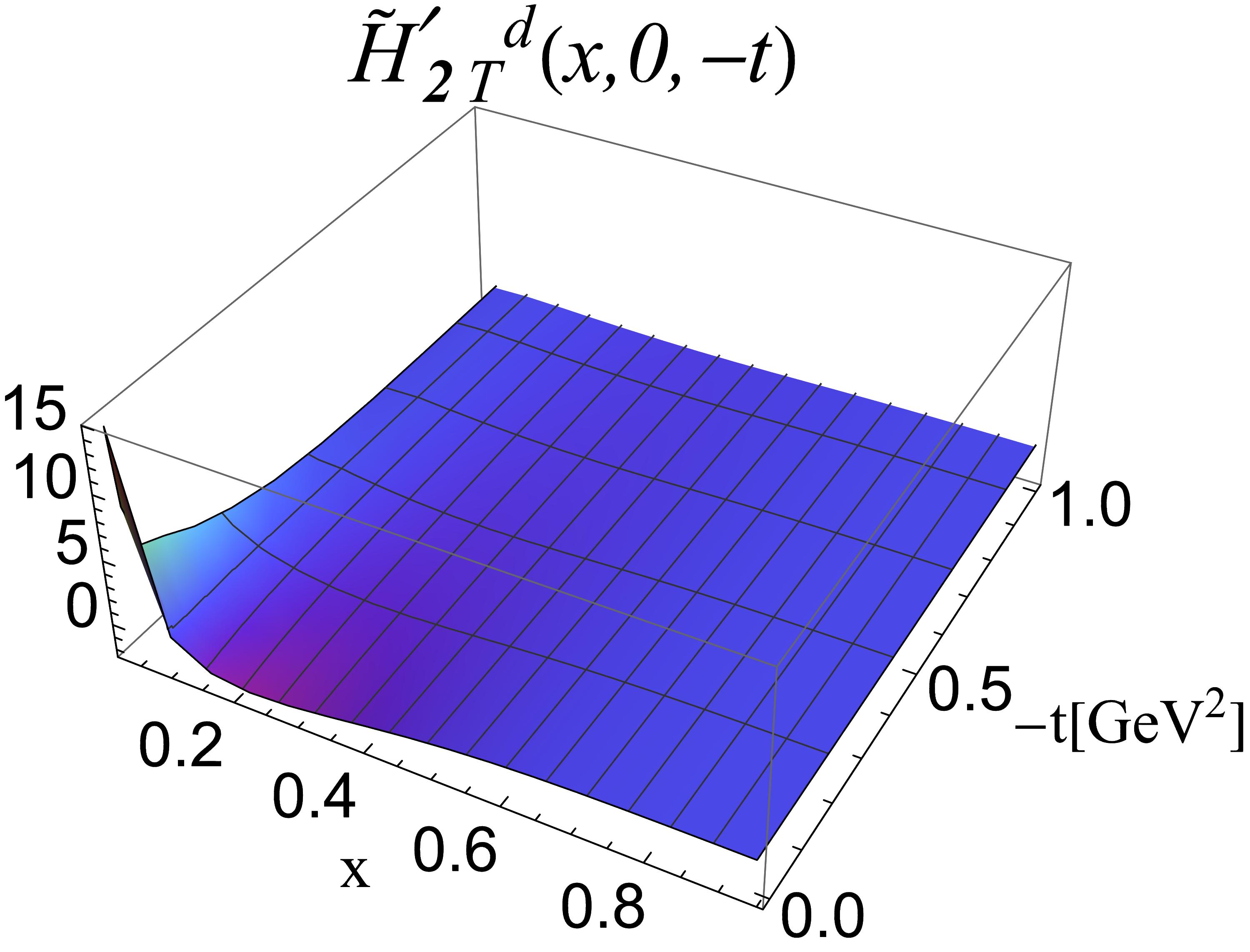}} \hspace{.3in}
    \subfloat[ ]{\includegraphics[scale=0.65]{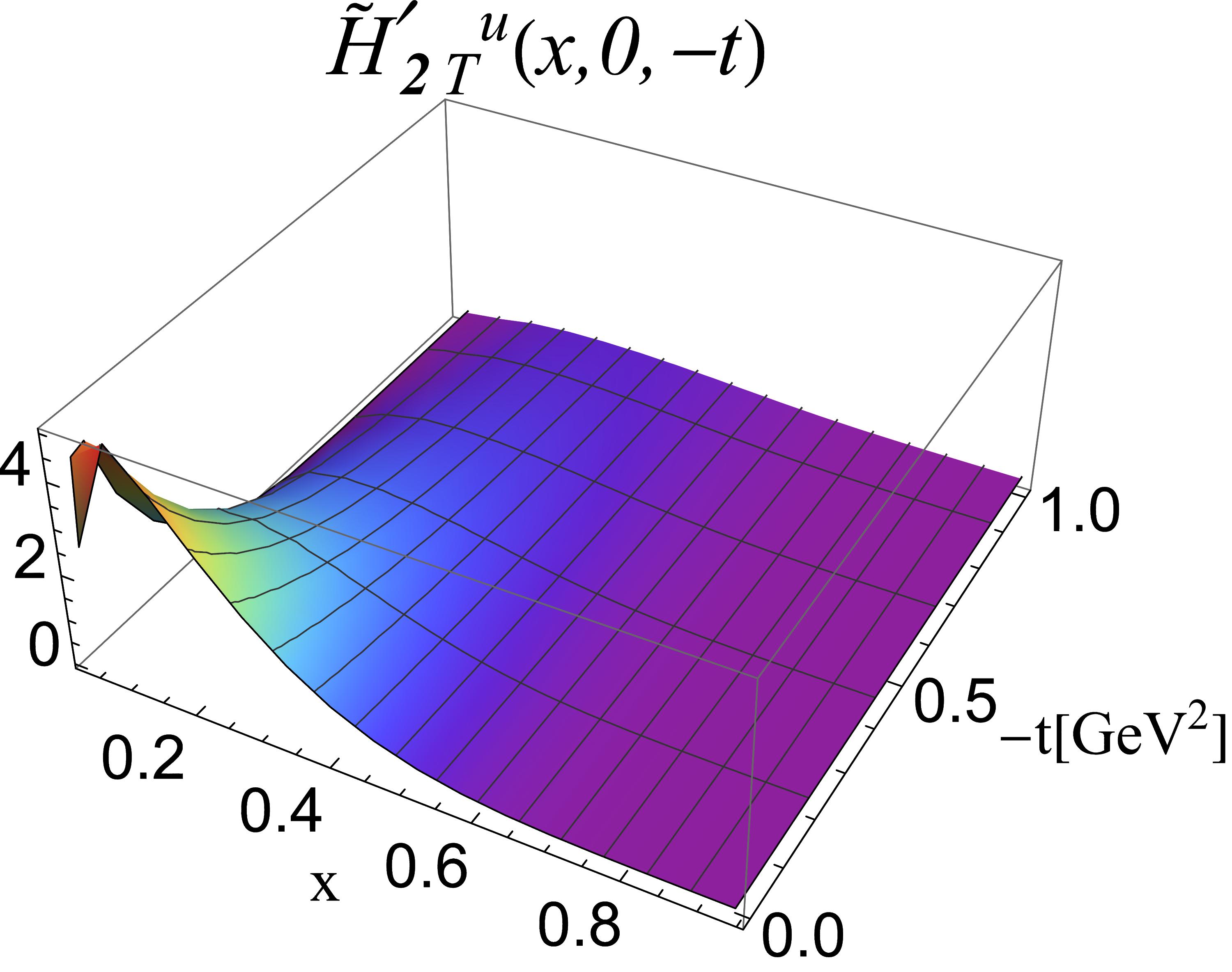}}
    \captionsetup{justification=raggedright}
    \caption{Twist-3 quark GPDs in the proton associated with $\Gamma=\gamma^{\perp} \gamma_{5}$; \{(a) (c) (e)\} and \{(b) (d) (f)\} are for the down and up quarks on the quark level, respectively. The flavor level distributions are given by $X_{flavor}^u =2 X_{quark}^u$ and $X_{flavor}^d=X_{quark}^d$, where $X$ stands for all the GPDs. The GPDs are evaluated with $N_{\mathrm{max}}=10$ and $K=16.5$.}
    \label{plotgammamugamma5}
\end{figure*}

\begin{figure*}[htbp]
    \centering
    \subfloat[ ]{\includegraphics[scale=0.65]{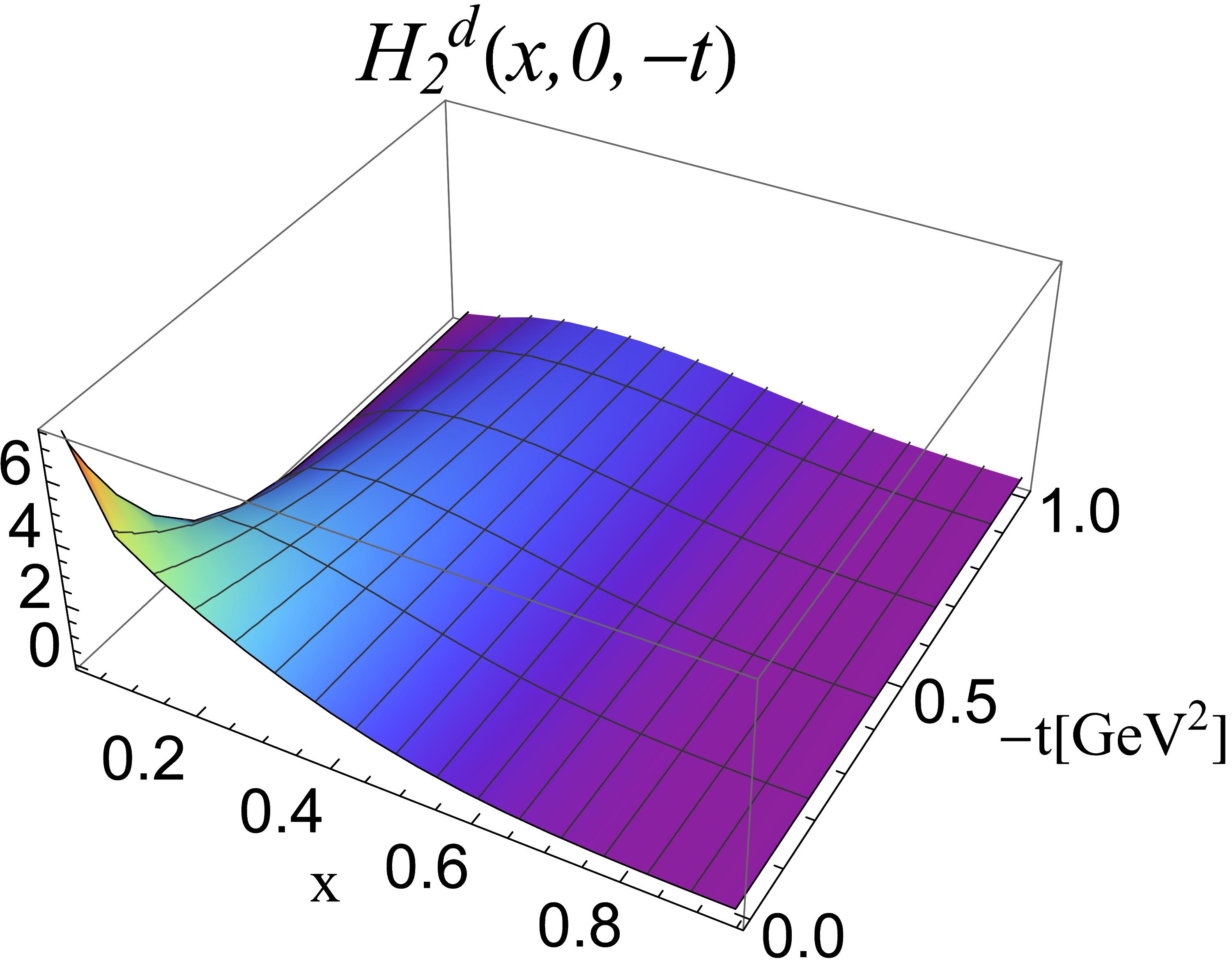}} \hspace{.3in}
    \subfloat[ ]{\includegraphics[scale=0.65]{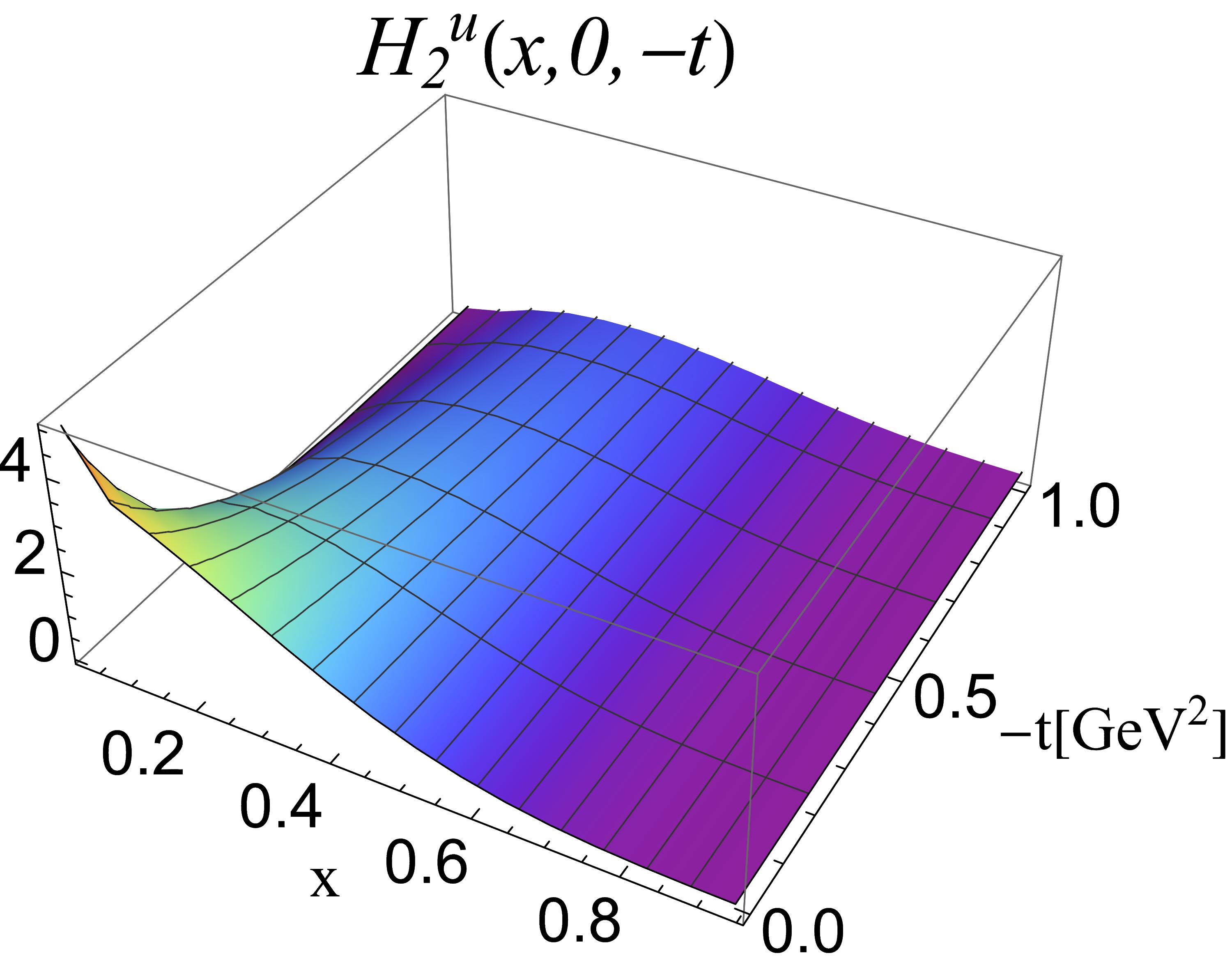}} \\
    \subfloat[ ]{\includegraphics[scale=0.65]{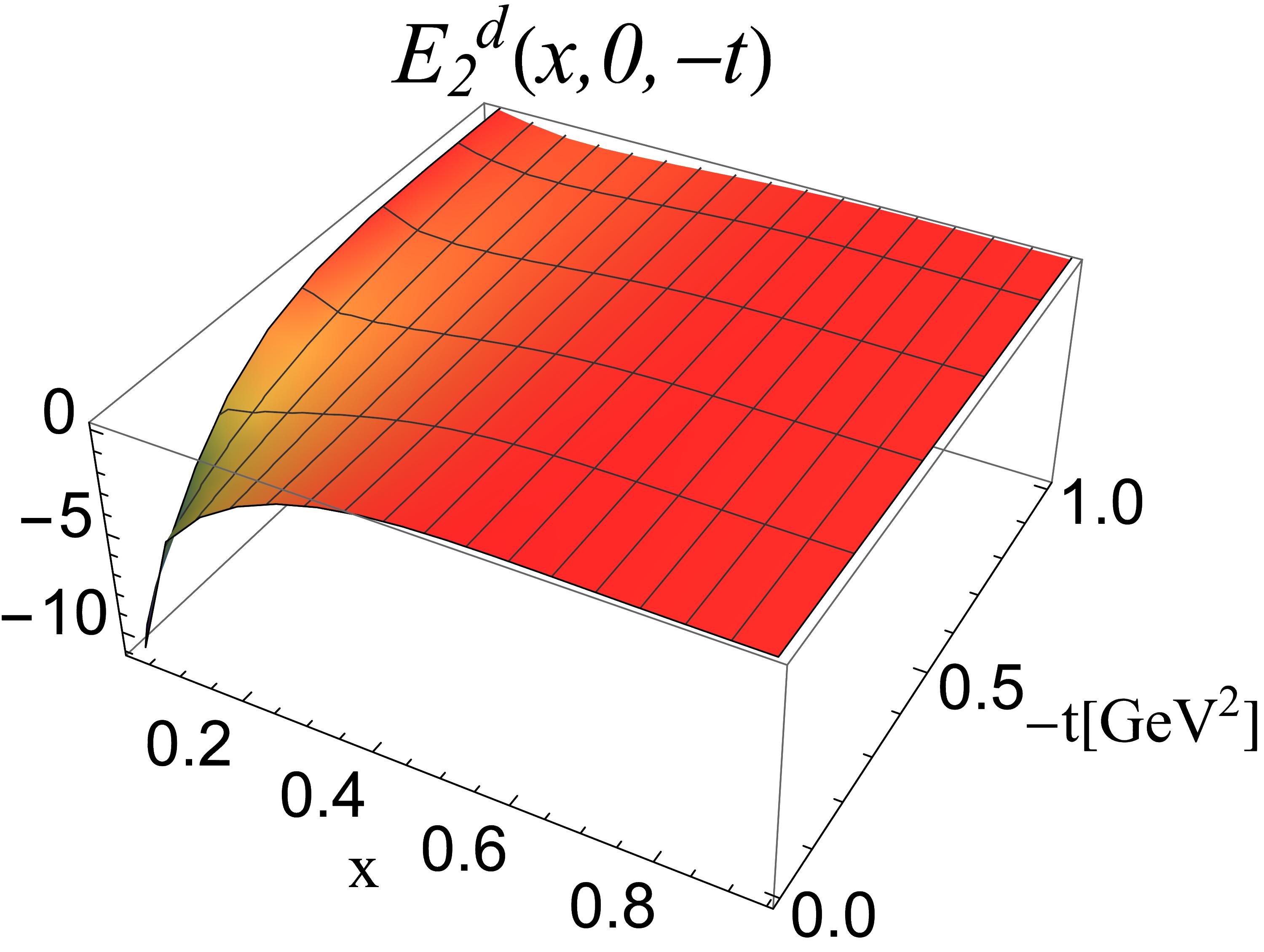}} \hspace{.3in}
    \subfloat[ ]{\includegraphics[scale=0.65]{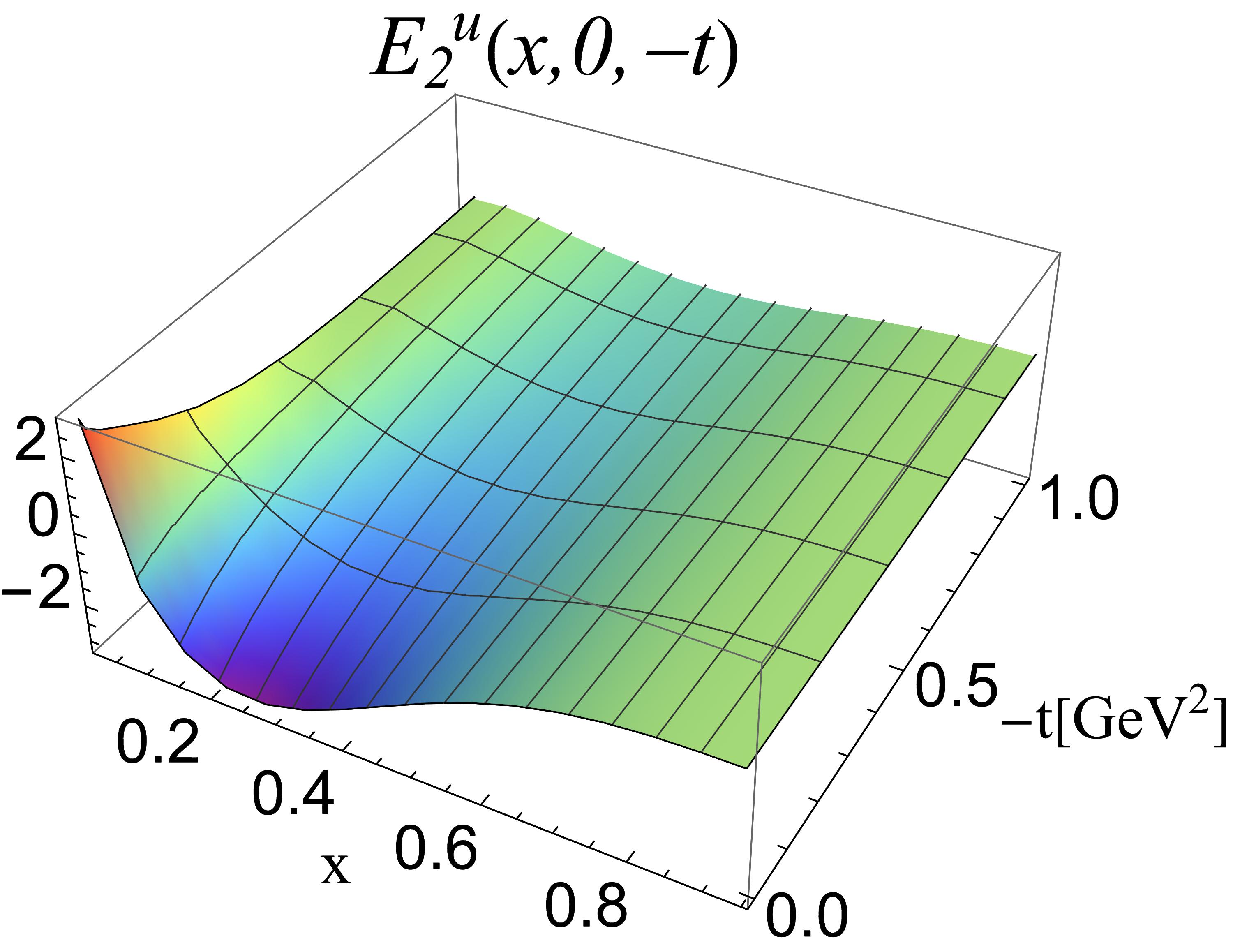}}
    \captionsetup{justification=raggedright}
    \caption{Twist-3 quark GPDs in the proton associated with $\Gamma=1$; \{(a) (c)\} and \{(b) (d)\} are for the down and up quarks on the quark level, respectively. The flavor level distributions are given by $X_{flavor}^u =2 X_{quark}^u$ and $X_{flavor}^d=X_{quark}^d$, where $X$ stands for all the GPDs. The GPDs are evaluated with $N_{\mathrm{max}}=10$ and $K=16.5$.}
    \label{plot1}
\end{figure*}

\begin{figure*}[htbp]
    \centering
    \subfloat[ ]{\includegraphics[scale=0.65]{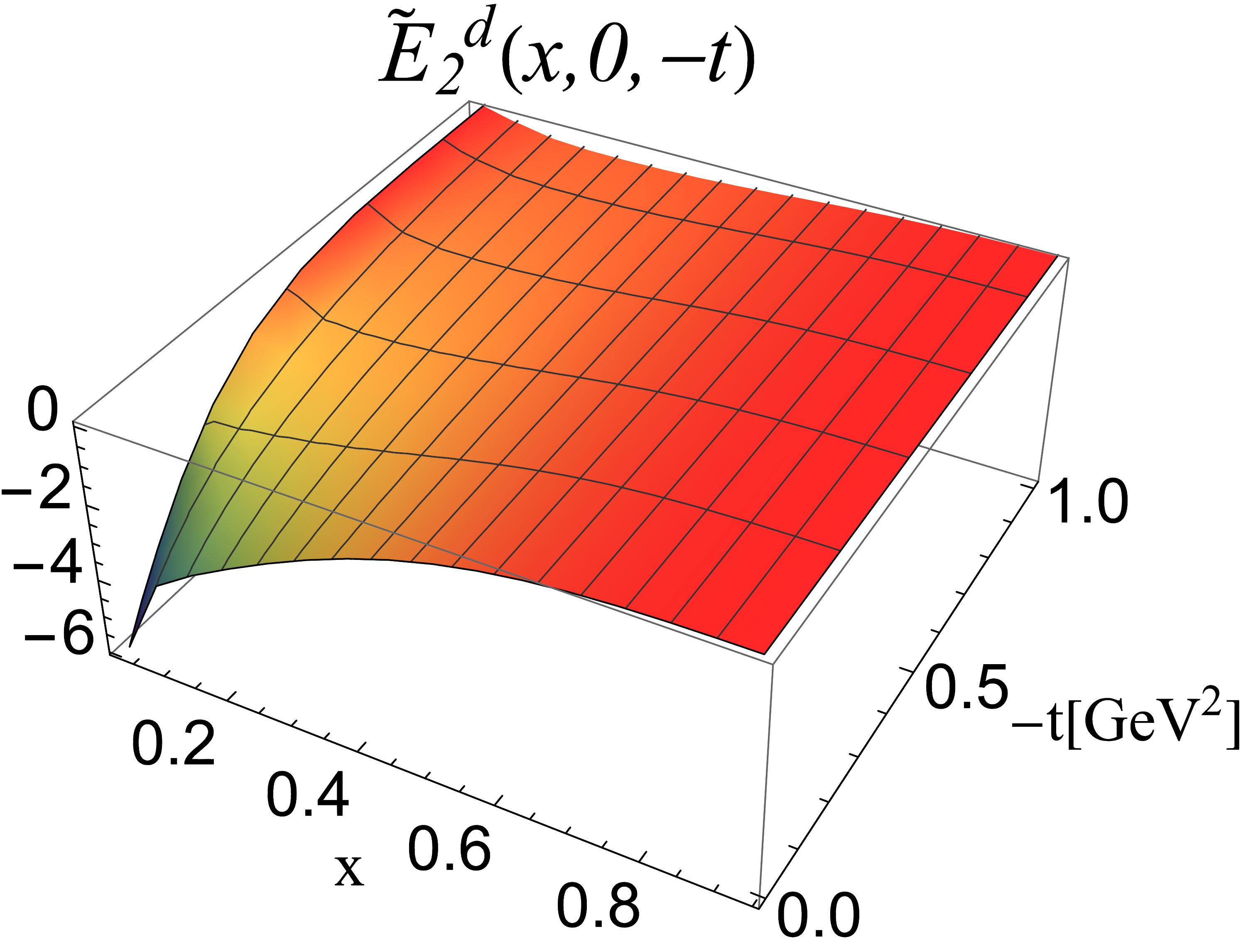}} \hspace{.3in}
    \subfloat[ ]{\includegraphics[scale=0.65]{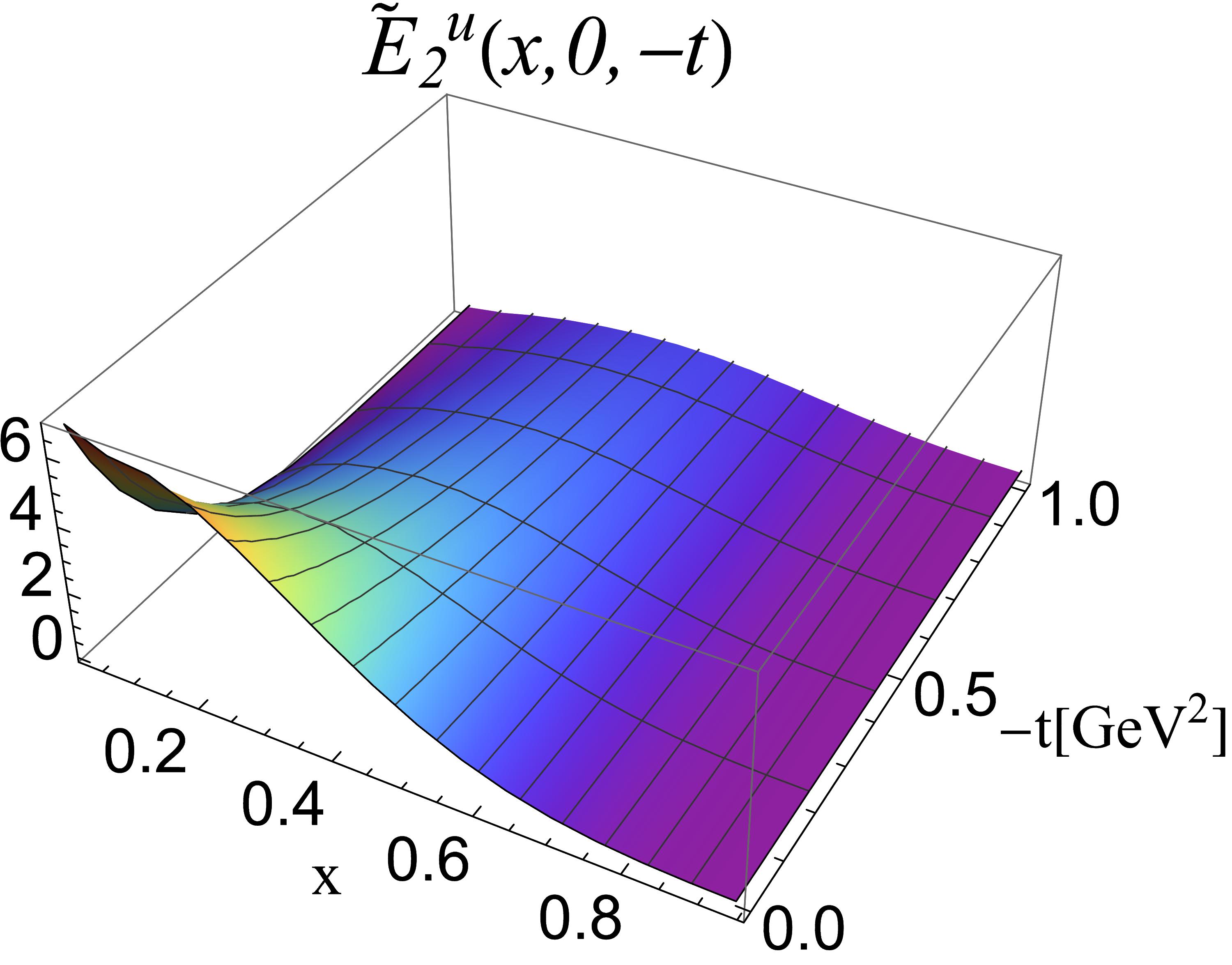}}
    \captionsetup{justification=raggedright}
    \caption{Twist-3 quark GPDs in the proton associated with $\Gamma=\gamma_{5}$; (a) and (b) are for the down and up quarks on the quark level, respectively. The flavor level distributions are given by $X_{flavor}^u =2 X_{quark}^u$ and $X_{flavor}^d=X_{quark}^d$, where $X$ stands for all the GPDs. The GPDs are evaluated with $N_{\mathrm{max}}=10$ and $K=16.5$.}
    \label{plotgamma5}
\end{figure*}

\begin{figure*}[htbp]
    \centering
    \subfloat[ test]{\includegraphics[scale=0.65]{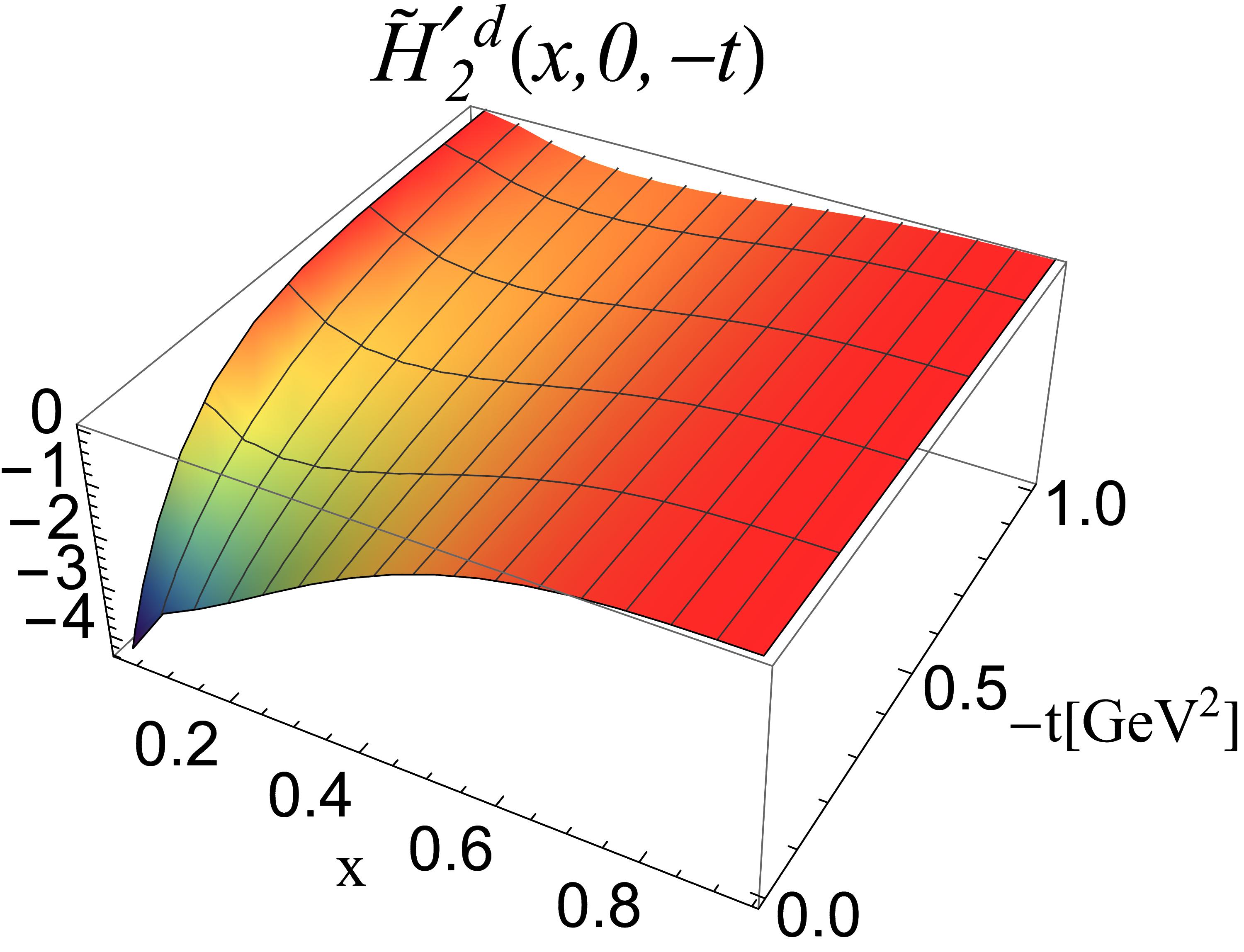}} \hspace{.3in}
    \subfloat[ ]{\includegraphics[scale=0.65]{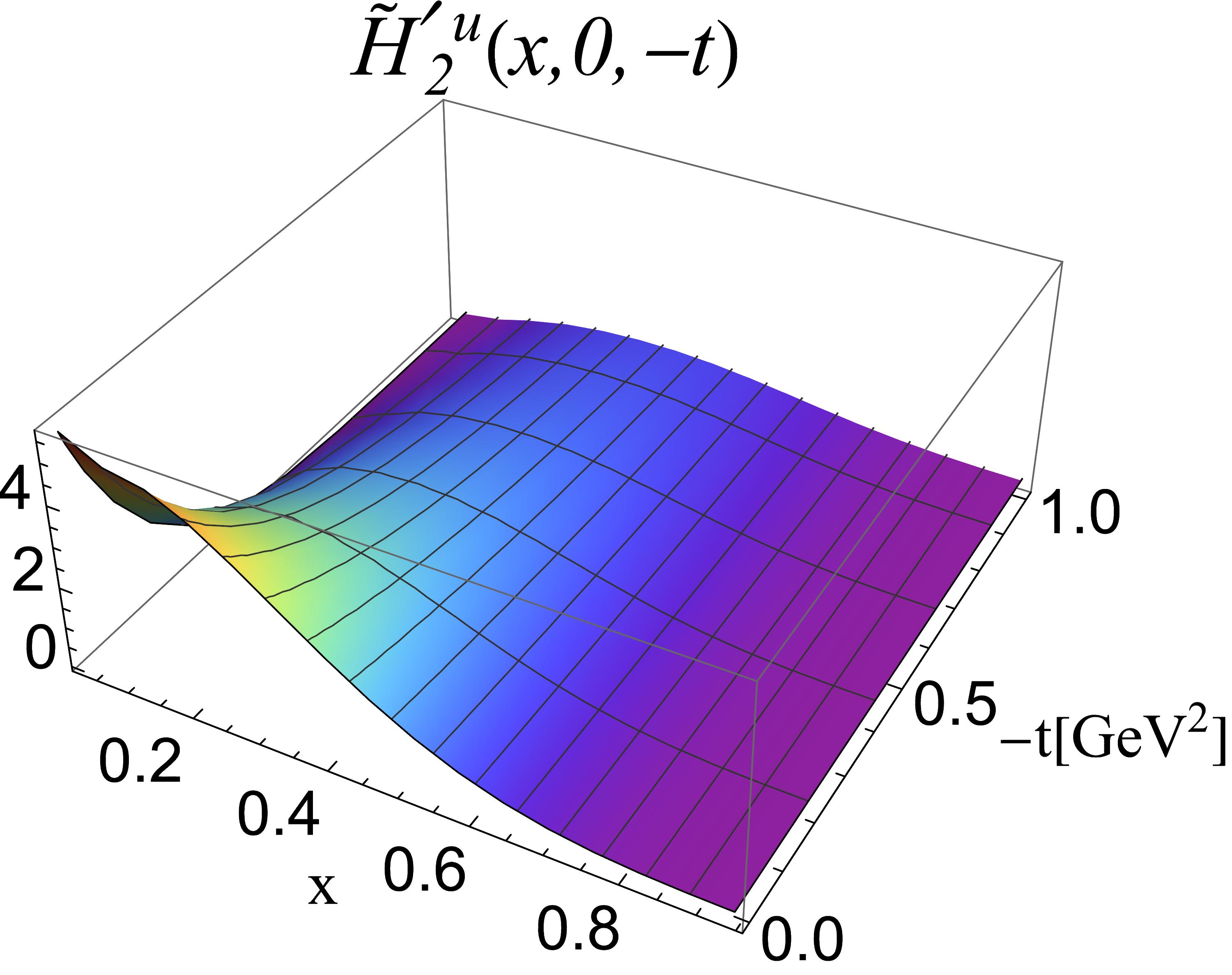}}
    \captionsetup{justification=raggedright}
    \caption{Twist-3 quark GPDs in the proton associated with $\Gamma=i \sigma^{+-} \gamma_5$; (a) and (b) are for the down and up quarks on the quark level, respectively. The flavor level distributions are given by $X_{flavor}^u =2 X_{quark}^u$ and $X_{flavor}^d=X_{quark}^d$, where $X$ stands for all the GPDs. The GPDs are evaluated with $N_{\mathrm{max}}=10$ and $K=16.5$.}
    \label{plotisigma}
\end{figure*}

With overlap representation and the LFWFs obtained from solving BLFQ with the specific implementations described above, we calculate all the twist-3 GPDs and display the results in Figs. (\ref{plotgammamu})-(\ref{plotisigma}). These overlap expressions in Eqs. (\ref{olpst})-(\ref{olf8ed}) are slightly different between $i=1$ and $i=2$, but we find that the numerical results are the same, which demonstrates rotational symmetry in the transverse plane within our approach. Thus we only show one of the two choices for both $u$ and $d$ quarks for simplicity. Since $\xi$ is fixed at zero in this work, we plot the GPDs as functions of $x$ and $-t$. In this zero-skewness limit, eight of the twist-3 GPDs, $H_{2T}$, $E_{2T}$, $\tilde{H}_{2T}$, $\tilde{E}'_{2T}$, $\tilde{H}_2$, $H'_2$, $E'_2$ and $\tilde{E}'_2$, are consistent with zero in our calculations to within our numerical uncertainty. This conclusion is consistent with their properties that we will discuss later.

$\tilde{E}_{2T}$ is the only $\gamma^{\perp}$ structure GPD that survives in the zero-skewness limit, and its 3D structures are shown in Fig. \ref{plotgammamu}. The distributions for both $d$ and $u$ quarks have peaks at small $x$ and small $-t$ region with opposite sign. This is due to the appearance of $x$ and $\Delta$ in the denominator of the expression of $\tilde{E}_{2T}$. The distribution of $\tilde{E}_{2T}$ falls very fast in the small $x$ region at small $-t$ followed by the appearance of a rounded peak moving to higher $x$ with increasing $-t$.

Now referring to Fig. \ref{plotgammamugamma5}, we first note that for the $\gamma^{\perp} \gamma_{5}$ related GPDs, only $\tilde{E}'_{2T}$ is zero. $H'_{2T}$ behaves like $\tilde{E}_{2T}$. $E'_{2T}$ possesses a dipole structure in the $x$ direction and, accordingly, has a peak in the small $-t$ region at around $x=0.2$ for both the $u$ and the $d$ quark as seen in Fig. \ref{plotgammamugamma5}. Similar structures appear in the $\tilde{H}'_{2T}$, where a dip is observed near the origin. We tested their structures with different parameter sets, and similar behavior persists. We suspect that this is a model-dependent feature related to the BLFQ calculations.

As depicted in Fig. \ref{plot1}, the behaviors of $H_2$ closely coincide for $u$ and $d$ quarks. However, for $E_{2}$, a distinct pattern emerges, exhibiting a peak near the origin for the $d$ quark and a distinct valley at around $x=0.2$ for the $u$ quark.

For $\tilde{E}_2$ and $\tilde{H}'_2$ in Figs. \ref{plotgamma5} and \ref{plotisigma}, they exhibit comparable trends with varying magnitudes. This alignment can be readily confirmed by examining the expressions for $\tilde{E}_{2}$ and $\tilde{H}'_{2}$ in Eqs. (\ref{eqgamma5}) and (\ref{eqend}).

\begin{figure*}[htbp]
    \centering
    \subfloat{\includegraphics[scale=0.62]{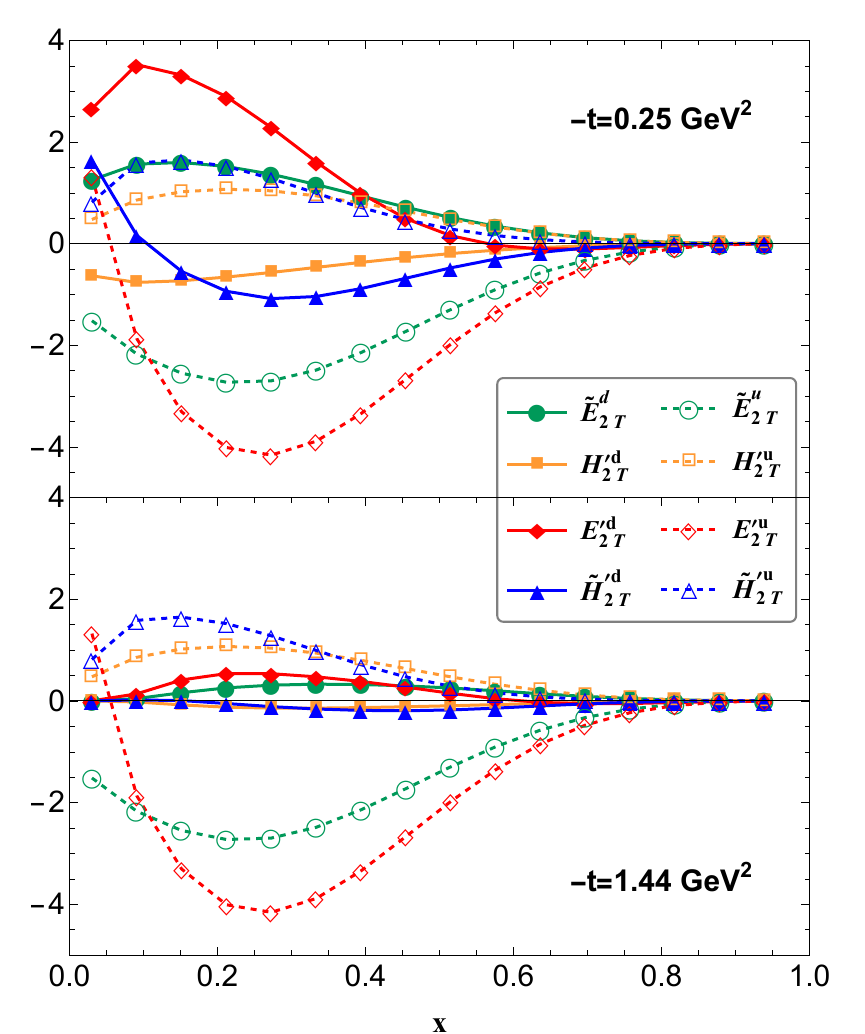}} 
    \subfloat{\includegraphics[scale=0.62]{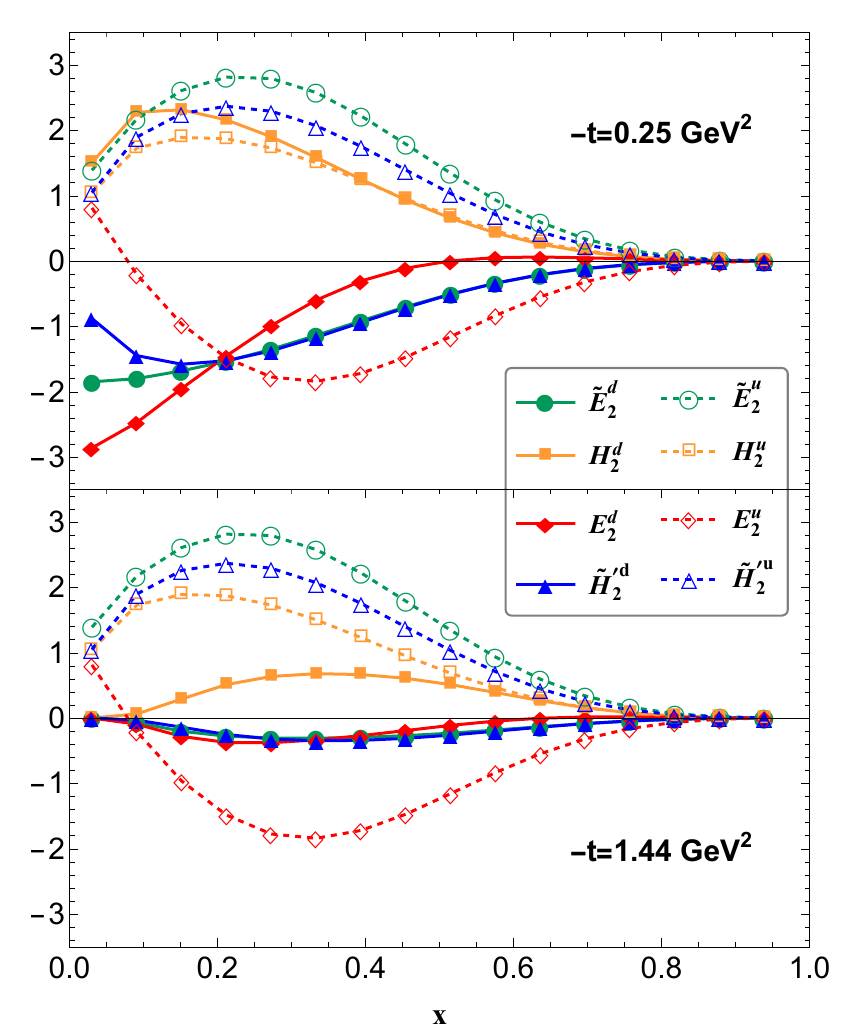}}
    \captionsetup{justification=raggedright}
    \caption{The upper row and the lower row are the twist-3 GPDs with fixed $-t=0.25\  \mathrm{GeV}^2$ and fixed $-t=1.44\  \mathrm{GeV}^2$ on the quark level respectively. The flavor level distributions are given by $X_{flavor}^u =2 X_{quark}^u$ and $X_{flavor}^d=X_{quark}^d$. We use the same color to represent the same GPD for each figure, and the solid lines with fill markers and the dashed lines with open markers to represent the down and up quarks respectively.} 
    \label{first2dx}
\end{figure*}

\begin{figure*}[htbp]
    \centering
    \subfloat{\includegraphics[scale=0.6]{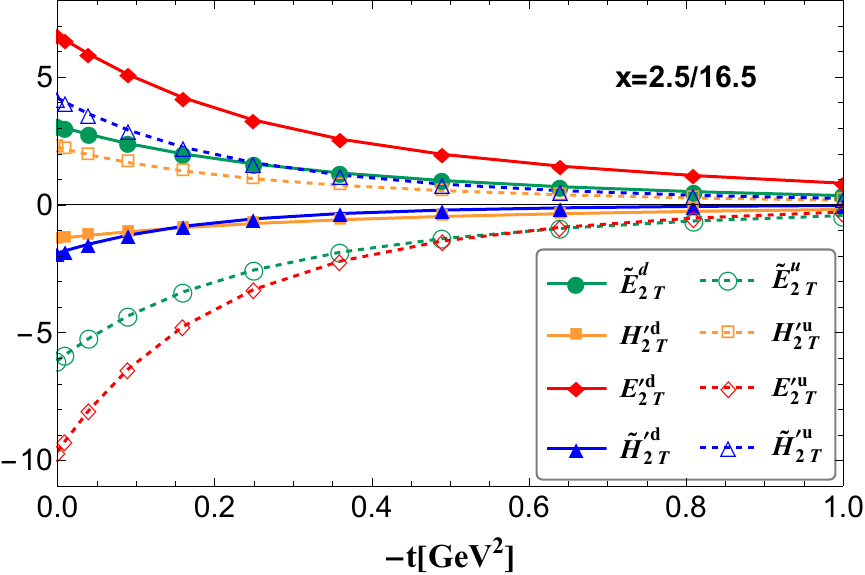}} \hspace{.15in}
    \subfloat{\includegraphics[scale=0.59]{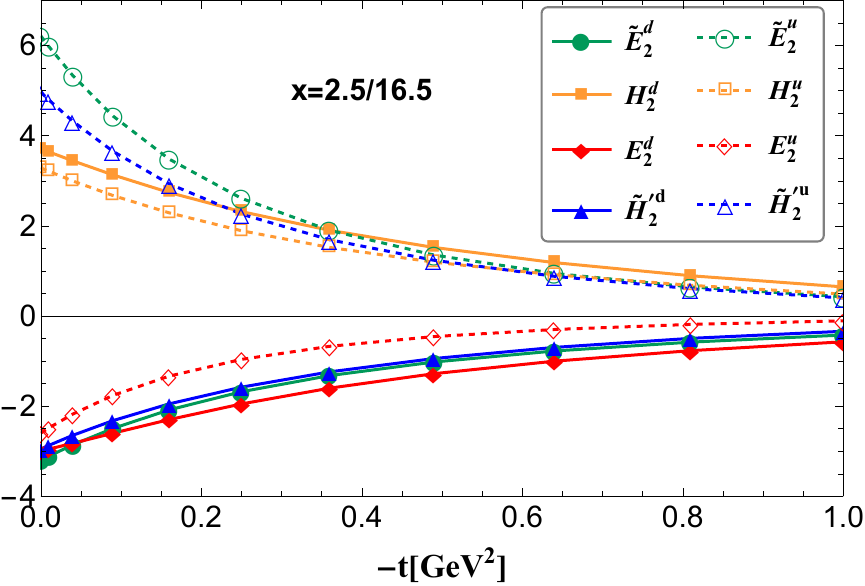}} 
    \captionsetup{justification=raggedright}
    \caption{The left figure and the right figure are the twist-3 GPDs with fixed $x=2.5/16.5$ on the quark level. The flavor level distributions are given by $X_{flavor}^u =2 X_{quark}^u$ and $X_{flavor}^d=X_{quark}^d$. We use one color to represent the same GPD for each figure, and the solid lines with fill markers and the dashed lines with open markers to represent the down and up quarks respectively.}
    \label{first2dt}
\end{figure*}

Key features of the GPDs pictured above are highlighted through 2D forms in Figs. \ref{first2dx} and \ref{first2dt}. From Fig. \ref{first2dx}, where $-t$ is fixed at $0.25 \ \mathrm{GeV}^2$ and $1.44 \ \mathrm{GeV}^2$, one finds that the peaks are moving towards higher $x$ as $-t$ increases. The 2D plots with fixed $x$ at $2.5/16.5$ as shown in Fig. \ref{first2dt}, display the smooth trend towards zero as $-t$ increases.

\begin{figure}[htbp]
    \centering
    \includegraphics[scale=0.6]{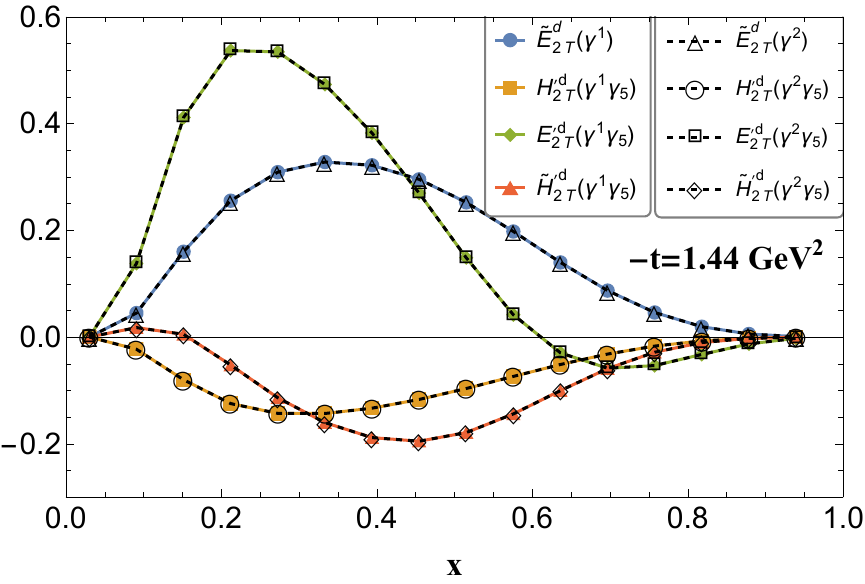}
    \captionsetup{justification=raggedright}
    \caption{The transverse symmetry of the twist-3 GPDs with $\gamma^{\perp}$, where we use colorful fill markers and black open markers to represent $\gamma^1$ and $\gamma^2$ related structure respectively.}
    \label{symmtrans}
\end{figure}

As mentioned before, the transverse symmetry between $\gamma^1$ and $\gamma^2$ has been verified in our calculations to within numerical precision. We choose $-t=1.44 \ \mathrm{GeV}^2$ for $\tilde{E}_{2T}$, $H'_{2T}$, $E'_{2T}$ and $\tilde{H}'_{2T}$ in Fig. \ref{symmtrans} to illustrate this conclusion. Clearly, $\gamma^1$ related distributions (solid lines) and $\gamma^2$ related distributions (dashed lines) coincide with each other very well.

\subsection{\label{sec:level2}Sum Rules}
Sum rules for the twist-3 GPDs $G_{i}$ and $\tilde{G}_{i}$ have been derived in Ref. \cite{kiptily2004genuine}. For $x^{0}$-moments, 
\begin{align}
    &\int^{1}_{-1} d x G_{i} (x,\xi,-t) = 0, \label{smrst} \\
    &\int^{1}_{-1} d x \tilde{G}_{i} (x,\xi,-t) = 0, \label{smr2} 
\end{align}
where $i=1,2,3,4$. For $x^{1}$-moments,
\begin{align}
    &\int^{1}_{-1} d x x G_{1} (x, \xi, -t) = \frac{1}{2} \frac{\partial}{\partial \xi} \int^{1}_{-1} d x x E(x, \xi, -t), \\
    &\int^{1}_{-1} d x x G_{2}(x, \xi, -t) \notag \\
    & =\frac{1}{2} \left[G_{A} \left(-t \right) - \int^{1}_{-1} d x x (H + E) (x, \xi, -t) \right], \label{oam} \\
    &\int^{1}_{-1} d x x G_{3}(x, \xi, -t)=0, \label{smr3} \\
    &\int^{1}_{-1} d x x G_{4}(x, \xi, -t)=0 ,\\
    &\int^{1}_{-1} d x x \tilde{G}_{1} (x, \xi, -t) \notag \\
    & =\frac{1}{2} \left[ F_{2} (-t) + (\xi \frac{\partial}{\partial \xi} - 1) \int^{1}_{-1} d x x \tilde{E} (x, \xi, -t) \right], \\
    &\int^{1}_{-1} d x x \tilde{G}_{2} (x, \xi, -t) \notag \\
    & =\frac{1}{2} \left[ \xi^{2} G_{E} \left(-t \right) - \frac{-t}{4M^{2}} F_{2}(-t) - \int^{1}_{-1} d x x \tilde{H} (x, \xi, -t) \right], \label{smr4} \\
    &\int^{1}_{-1} d x x \tilde{G}_{3} (x, \xi, -t) = \frac{1}{4} \xi G_{E} (-t), \label{smr5} \\
    &\int^{1}_{-1} d x x \tilde{G}_{4} (x, \xi, -t) = \frac{1}{4} G_{E} (-t) \label{smred} ,
\end{align}
where $F_{1}(-t)$ and $F_{2}(-t)$ are the Dirac and the Pauli form factors, $G_{P}(-t)$ and $G_{A}(-t)$ are the pseudo-scalar and the axial-vector form factors, $G_{M}(-t)$ and $G_{E}(-t)$ are the magnetic and the electric form factors respectively. From definitions, we have
\begin{align}
    & \int^{1}_{-1} d x H (x,\xi=0,-t) = F_{1} (-t), \\
    & \int^{1}_{-1} d x E (x,\xi=0,-t) = F_{2} (-t), \\
    & \int^{1}_{-1} d x \tilde{H} (x,\xi=0,-t) = G_{A} (-t), \\
    & \int^{1}_{-1} d x \tilde{E} (x,\xi=0,-t) = G_{P} (-t), \\
    & G_{M}(-t) = F_{1}(-t) + F_{2}(-t), \\
    & G_{E}(-t) = F_{1}(-t) + \frac{-t}{4M^{2}} F_{2}(-t).
\end{align}

Note that opposite sign of the right hand side (r.h.s) of Eq. (\ref{oam}) after taking forward limit is exactly the so-called kinetic OAM defined in Ref. \cite{ji1997gauge}. This implies that twist-3 GPDs also have physical relevance that is not negligible.

Using the inversion relations, Eqs. (\ref{invst})-(\ref{inved}), one can express the sum rules, Eqs. (\ref{smrst})-(\ref{smred}) in terms of GPDs in Eqs. (\ref{gammagpd}) and (\ref{gamma5gpd}). This work mainly focuses on the zero-skewness limit, $\xi = 0$, so some of the sum rules above, which contain $1/ \xi$ or $\partial / \partial \xi$ will not be discussed here. The twist-2 GPDs used here have been calculated with the same parameter set and truncation \cite{Xu:2021wwj,Liu:2022fvl}. Note that one has to consider nonzero skewness, $\xi\ne0$ to compute the GPD $\tilde{E}$. Using the properties of GPDs under discrete symmetries (see Appendix \ref{dissym}), some of the GPDs are already zero at both the theoretical and the numerical level, making the sum rules (\ref{smrst}) ($i=1,3$), (\ref{smr2}) ($i=3$) and (\ref{smr5}) automatically satisfied. The differences between the left hand side (l.h.s) and the r.h.s of Eqs. (\ref{smr2}) ($i=2,4$) (\ref{smr4}) and (\ref{smred}) are not small enough to be treated as the numerical error. However, they all exhibit a common trend that they decrease as $-t$ increases. We suspect that the deviation is due to the fact that we have only retained the leading Fock-sector and will discuss those results in more details later when we extend to the higher Fock-sectors to include contributions arising from a dynamical gluon.

In Ref. \cite{guo2021novel}, the authors investigate the relation between the GPDs in Eq. (\ref{par2}) and the gravitational form factors (GFFs) as the parameterization of the matrix elements of the energy momentum tensor (EMT). Following some constraints of the GFFs, the following sum rules are obtained
\begin{equation}
    \int dx x \lim_{\Delta \rightarrow 0} G_{q,j}(x,\xi,-t) = 0, \label{emtff}
\end{equation}
with $j={1,2,4}$. With the replacement of Eqs. (\ref{jis1}), (\ref{jis2}) and (\ref{jis4}), we find Eq. (\ref{emtff}) is satisfied each time to within numerical precision.

\subsection{\label{sec:level2}PDF Limit}
The PDF was originally introduced by Feynman to describe the deep inelastic lepton scattering process, providing information on the hadron structure in the longitudinal direction. The PDFs can be interpreted as parton densities corresponding to a specified longitudinal momentum fraction $x$. The PDFs are also defined by a quark-quark density matrix \cite{jaffe1992chiral},
\begin{equation}
    \mathscr{F}(x) = \int \frac{d y^-}{2\pi} e^{i y^- x} \langle P,\Lambda | \bar{\psi} (-\frac{y^-}{2}) \Gamma \psi (\frac{y^-}{2}) | P,\Lambda \rangle,
\end{equation}

By taking different $\Gamma$ structures, we could get the PDFs for all twist. This work only focuses on twist-3 PDFs, which are given by
\begin{align}
    \int \frac{d y^-}{2\pi} e^{i y^- x} \langle P,\Lambda | \bar{\psi}(-\frac{y}{2}) \psi(\frac{y}{2}) | P,\Lambda \rangle &= 2M e(x), \\
    \int \frac{d y^-}{2\pi} e^{i y^- x} \langle P,\Lambda | \bar{\psi}(-\frac{y}{2}) \gamma^{i} \gamma_{5} \psi(\frac{y}{2}) | P,\Lambda \rangle &= 2 g_{T}(x), \label{gt} \\
    \int \frac{d y^-}{2\pi} e^{i y^- x} \langle P,\Lambda | \bar{\psi}(-\frac{y}{2}) i \sigma^{+-} \gamma_{5} \psi(\frac{y}{2}) | P,\Lambda \rangle &= 2 h_{L}(x) M, \label{hl}
\end{align}
where the indice $i$ in Eq. (\ref{gt}) can only be $1$ or $2$. 

\begin{figure}[htbp]
    \centering
    \includegraphics[scale=0.58]{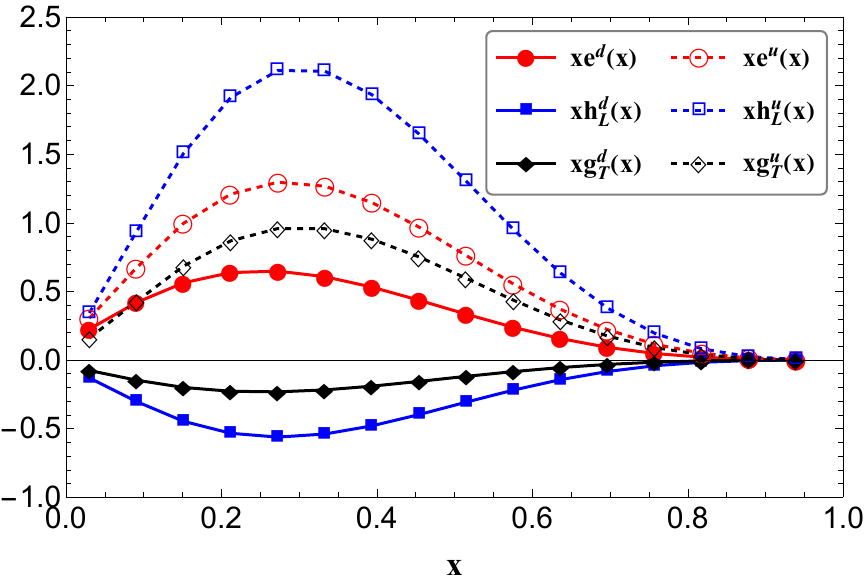}
    \captionsetup{justification=raggedright}
    \caption{Twist-3 PDFs in the proton on the flavor level with different colour curves; the fill markers and the open markers are for the down and up quarks, respectively. The PDFs are evaluated with $N_{\mathrm{max}}=10$ and $K=16.5$.}
    \label{plotpdf}
\end{figure}

It can be immediately found that some pairs of the GPDs and the PDFs are connected to each other by taking the forward limit
\begin{align}
    e(x) = \lim \limits_{\Delta \rightarrow 0} H_2(x,0,-t), \label{exdef} \\
    g_{T}(x) = \lim \limits_{\Delta \rightarrow 0} H'_{2T}(x,0,-t), \label{gtdef} \\
    h_{L}(x) = \lim \limits_{\Delta \rightarrow 0} \tilde{H}'_2(x,0,-t). \label{hldef}
\end{align}
Since $\Delta$ appears in the denominator of the expressions of $H'_{2T}$, we try to perform the limit numerically to obtain the twist-3 PDFs. Fortunately, we find that our results are numerically converging when $-t$ is smaller than $10^{-7} \ \mathrm{GeV}^{2}$. So we choose $-t=10^{-20} \ \mathrm{GeV}^{2}$ as the numerical point representing the forward limit, and the results are shown in Fig. \ref{plotpdf}.

Our results satisfy the sum rule \cite{jaffe1992chiral}
\begin{equation}
    \int x e(x) dx = \frac{m_{q}}{M} N_{q},
\end{equation}
where $m_{q}$ is the quark mass and $N_{q}$ is the number of valence quarks of flavor $q$.

More interesting things happen to $g_{T}(x)$. At twist-2 level, the similar gamma structure parameterised as $g_{1}(x)$ is the so-called helicity PDF that measures the quark helicity distribution in a longitudinally polarized nucleon. At the twist-3 level, Eqs. (\ref{gtdef}) and (\ref{gth2tp}) show that $g_{T}(x)$ measures the transverse spin distribution for quarks. Here there is also a well known sum rule called the Burkhardt-Cottingham sum rule \cite{Burkhardt:1970ti},
\begin{equation}
    \int g_{2}(x) dx = 0,
\end{equation}
where $g_{2}(x) = g_{T}(x) - g_{1}(x)$. This sum rule means that the contributions of quark spin to the proton spin with different polarizations are the same. But this sum rule is derived from three-dimensional rotational symmetry, which is broken by the truncation of Fock sector and is therefore not satisfied in this work. We hope that with the inclusion of higher Fock sectors in the future this will be improved. As for $h_{L}(x)$, it contributes to a polarized Drell-Yan process. Both $g_{T}(x)$ and $h_{L}(x)$ are sensitive to quark-gluon interaction \cite{jaffe1992chiral}.

Recently, the point-by-point extraction of the twist-3 scalar PDF $e(x)$ through the analysis of both CLAS and CLAS12 data for dihadron production in semi-inclusive DIS off of a proton target has been reported in Ref. \cite{courtoy2022extraction}. We also notice that there are many model calculations of $e(x)$ for which we provide a comparison below. The proton flavor combination quantity $e_{v}$ is defined by \cite{courtoy2022extraction,pasquini2019twist,Lorce:2014hxa}
\begin{equation}
    e_{v} = \frac{4}{9} (e_{u} - e_{\bar{u}}) - \frac{1}{9} (e_{d} - e_{\bar{d}}).
\end{equation}

\begin{figure}[htbp]
    \centering
    \includegraphics[scale=0.6]{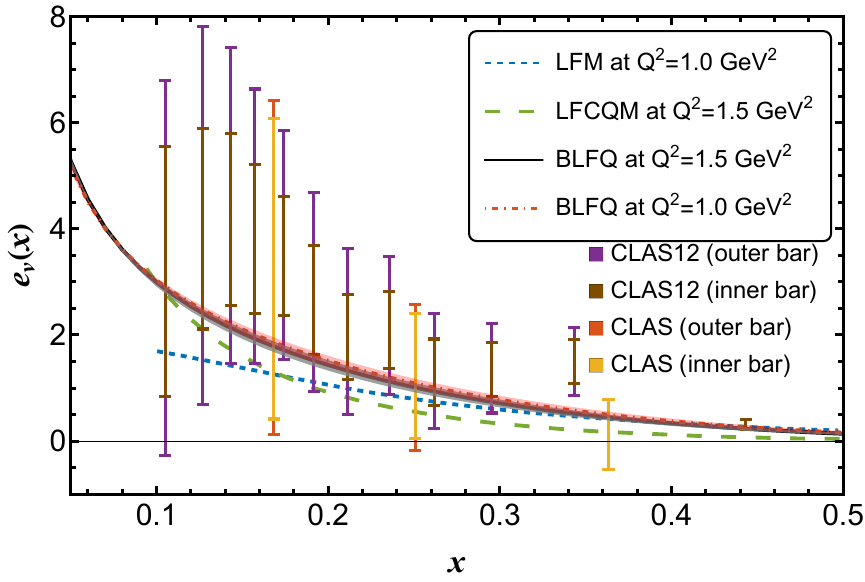}
    \captionsetup{justification=raggedright}
    \caption{BLFQ calculations for proton-flavor combination $e_{v}$ at $Q^{2}=1.0 \mathrm{GeV}^{2}$ (red line) and $Q^{2}=1.5 \mathrm{GeV}^{2}$ (black line) in comparison with other results. The blue line is extracted from Ref. \cite{pasquini2019twist}, the green line is extracted from Ref. \cite{Lorce:2014hxa}, the purple and brown error bars are the extraction of CLAS data, and the red and yellow error bars are the extraction of CLAS12 data. Here the inner bars represent the contribution from the 0th approximation, corrected by the twist-3 contributions for the fragmentation sector for the outer bars as explained in \cite{courtoy2022extraction}.}
    \label{evloved}
\end{figure}
In Fig. \ref{evloved}, we present the $e_{v}$ for both theoretical model calculations and experimental extractions. We utilize the Higher Order Perturbative Parton Evolution toolkit (HOPPET) to numerically solve the NNLO DGLAP equations \cite{salam2009higher}, and evolve BLFQ results from the initial scale $\mu^{2}_{0} = 0.195 \pm 0.02 \ \mathrm{GeV}^{2}$ \cite{Xu:2021wwj} to the scale at $Q^{2}=1.0 \ \mathrm{GeV}^{2}$ and $Q^{2}=1.5 \ \mathrm{GeV}^{2}$. The error bands in our evolved distributions are due to the $10\%$ uncertainties in the initial
scale. We find that our results show good agreement with other model calculations and the experimental data in most regions. In the small $x$ region the experimental data rise sharply, whereas our results rise more slowly. It should also be noted that there is large uncertainty in the experimental data in the low $x$ region. Despite that, our preliminary results are generally consistent with most of the model calculations and the experimental data.

\section{\label{sec:level1}Conclusion} \label{sec5}
In this paper, we present all the twist-3 GPDs in the zero-skewness limit of the proton within the theoretical framework of Basis Light-front Quantization (BLFQ). The LFWFs are obtained by diagonalizing the proton light-front Hamiltonian using BLFQ. Then the LFWFs are used to calculate twist-3 GPDs via the overlap representation. We have calculated all well-defined twist-3 GPDs, both chiral-odd and chiral-even, with a truncation to leading Fock sector and numerical cut offs. We also calculate twist-3 PDFs and evolve the spin-independent PDF $e(x)$ to a higher scale that we then compare with other theoretical model calculations and with experimental data. We find there is general agreement among the selected models and rough agreement between models and experiment considering the large experimental uncertainties.

In the future, we shall calculate twist-3 GPDs with higher Fock sectors, and study the relation between twist-3 GPDs and orbital angular momentum distribution. The study of nonzero-skewness is also of interest because it can be connected to DVCS twist-3 cross section \cite{guo2022twist} and can be measured through the experiments at the EIC and EicC \cite{chavez2022accessing}. Results at twist-3 and non-zero skewness will deepen our understanding of the proton structure, and further, help to solve the proton spin puzzle.

\begin{acknowledgments}
We thank Jiangshan Lan and Jiatong Wu for many helpful discussions and useful advises. C. M. is supported by new faculty start up funding the Institute of Modern Physics, Chinese Academy of Sciences, Grants No. E129952YR0. C. M. also thanks the Chinese Academy of Sciences Presidents International Fellowship Initiative for the support via Grants No. 2021PM0023. X. Z. is supported by new faculty startup funding by the Institute of Modern Physics, Chinese Academy of Sciences, by Key Research Program of Frontier Sciences, Chinese Academy of Sciences, Grant No. ZDB-SLY-7020, by the Natural Science Foundation of Gansu Province, China, Grant No. 20JR10RA067, by the Foundation for Key Talents of Gansu Province, by the Central Funds Guiding the Local Science and Technology Development of Gansu Province, Grant No. 22ZY1QA006, by international partnership program of the Chinese Academy of Sciences, Grant No. 016GJHZ2022103FN, by the National Natural Science Foundation of China under Grant No. 12375143, by National Key R\&D Program of China, Grant No. 2023YFA1606903, and by the Strategic Priority Research Program of the Chinese Academy of Sciences, Grant No. XDB34000000. J. P. V. is supported by the US Department of Energy under Grant Nos. DE-SC0023692 and DE-SC0023707. A portion of the computational resources were also provided by Gansu Computing Center. This research is supported by Gansu International Collaboration and Talents Recruitment Base of Particle Physics (2023-2027) and the International Partnership Program of Chinese Academy of Sciences, Grant No.016GJHZ2022103FN.
\end{acknowledgments}

\appendix

\section{Relations Between Different GPDs} \label{app}
The relations between GPDs defined in Ref. \cite{PhysRevD.98.014038} and those in Eqs. (\ref{gammagpd})-(\ref{eqend}) are
\begin{align}
    H_{2T} =& 2\xi G_{4}, \\
    E_{2T} =& 2(G_{3} - \xi G_{4}), \\
    \tilde{H}_{2T} =& \frac{1}{2} G_{1}, \\
    \tilde{E}_{2T} =& -(H + E + G_{2}) + 2(\xi G_{3} - G_{4}), \\
    H'_{2T} =& \frac{-t}{4M^{2}} (\tilde{E} + \tilde{G}_{1}) + (\tilde{H} + \tilde{G}_{2}) - 2\xi \tilde{G}_{3}, \\
    E'_{2T} =& -(\tilde{E} + \tilde{G}_{1}) - (\tilde{H} + \tilde{G}_{2}) + 2(\xi \tilde{G}_{3} - \tilde{G}_{4}), \\
    \tilde{H}'_{2T} =& \frac{1}{2} (\tilde{E} + \tilde{G}_{1}), \\
    \tilde{E}'_{2T} =& 2(\tilde{G}_{3} - \xi \tilde{G}_{4}) .
\end{align}

The inversion of above equations are
\begin{align}
    G_{1} =& 2\tilde{H}_{2T}, \label{invst} \\
    G_{2} =& -(H + E) - \frac{1}{\xi} (1-\xi^{2}) H_{2T} + \xi E_{2T} - \tilde{E}_{2T}, \\
    G_{3} =& \frac{1}{2} (H_{2T} + E_{2T}), \\
    G_{4} =& \frac{1}{2\xi} H_{2T}, \\
    \tilde{G}_{1} =& -\tilde{E} + 2\tilde{H}'_{2T}, \\
    \tilde{G}_{2} =& -\tilde{H} + (1-\xi^{2}) H'_{2T} -\xi^{2} E'_{2T} - \frac{\Delta^{2}_{\perp}}{2M^{2}} \tilde{H}'_{2T} + \xi \tilde{E}'_{2T}, \\
    \tilde{G}_{3} =& -\frac{\xi}{2} (H'_{2T} + E'_{2T}) - \frac{\xi \bar{M}^{2}}{M^{2}} \tilde{H}'_{2T} + \frac{1}{2} \tilde{E}'_{2T}, \\
    \tilde{G}_{4} =& -\frac{1}{2} (H'_{2T} + E'_{2T}) - \frac{\bar{M}^{2}}{M^{2}} \tilde{H}'_{2T} \label{inved} .
\end{align}
where $\bar{M}^{2} = M^{2} + t/4$.

The relations between GPDs defined in Eq. (\ref{par2}) and this work are 
\begin{align}
    E_{2T} (x,\xi,-t) &= 2 G_{q,2} (x,\xi,-t), \label{jis1} \\
    H_{2T} (x,\xi,-t) &= G_{q,4} (x,\xi,-t), \label{jis2} \\
    \tilde{E}_{2T} (x,\xi,-t) &= 2\xi G_{q,2} (x,\xi,-t) - G_{q,3} (x,\xi,-t), \\
    \tilde{H}_{2T} (x,\xi,-t) &= G_{q,1} (x,\xi,-t). \label{jis4}
\end{align}

\section{\label{sec:level2}Discrete Symmetry} \label{dissym}
We address the properties of GPDs under discrete symmetries. Refs. \cite{meissner2009generalized,diehl2001generalized,Belitsky2005UnravelingHS} have shown that time-reversal and Hermitian conjugation together provide some interesting results. Here we only give a brief discussion. The time-reversal operator $\mathscr{T}$ is defined in the Hilbert space is an anti-unitary operator, and it has the following properties for any c-function $f$
\begin{align}
    \mathscr{T} f =& f^{\star} \mathscr{T} , \\
    \mathscr{T} (f_1 + f_2) =& \mathscr{T} f_1 + \mathscr{T} f_2 .
\end{align}
and for the inner product of quantum states
\begin{equation}
    \langle \psi_1 | \overleftarrow{\mathscr{T}} \overrightarrow{\mathscr{T}} | \psi_2 \rangle =  \langle \psi_1 | \psi_2 \rangle^{\star} = \langle \psi_2 | \psi_1 \rangle.
\end{equation}

If we act with them on Eqs. (\ref{gammagpd})-(\ref{eqend}) then some symmetries are obtained
\begin{equation}\label{plusnonconj}
    F^q(x,\xi,\Delta^2) = F^q(x,-\xi,\Delta^2),
\end{equation}
for $F=\tilde{E}_{2T},H'_{2T},E'_{2T},\tilde{H}'_{2T},\tilde{E}_2,H_2,E_2,\tilde{H}'_2$, and 
\begin{equation}\label{minusnonconj}
    F^q(x,\xi,\Delta^2) = -F^q(x,-\xi,\Delta^2).
\end{equation}
for $F=H_{2T},E_{2T},\tilde{H}_{2T},\tilde{E}'_{2T},\tilde{H}_2,H'_2,E'_2,\tilde{E}'_2$.

Taking the Hermitian conjugation on Eqs. (14-19) gives
\begin{equation}\label{plusconj}
    F^{q\star}(x,\xi,\Delta^2) = F^q(x,-\xi,\Delta^2),
\end{equation}
for $F=\tilde{E}_{2T},H'_{2T},E'_{2T},\tilde{H}'_{2T},\tilde{E}_2,H_2,E_2,\tilde{H}'_2$, and 
\begin{equation}\label{minusconj}
    F^{q\star}(x,\xi,\Delta^2) = -F^q(x,-\xi,\Delta^2).
\end{equation}
for $F=H_{2T},E_{2T},\tilde{H}_{2T},\tilde{E}'_{2T},\tilde{H}_2,H'_2,E'_2,\tilde{E}'_2$.

From Eq. (\ref{plusnonconj}) one finds that $\tilde{E}_{2T}$, $H'_{2T}$, $E'_{2T}$, $\tilde{H}'_{2T}$, $\tilde{E}_2$, $H_2$, $E_2$, $\tilde{H}'_2$ are even functions of $\xi$, while those in Eq. (\ref{minusnonconj}) are odd functions. This means that in the zero-skewness ($\xi=0$) limit, $H_{2T}$, $E_{2T}$, $\tilde{H}_{2T}$, $\tilde{E}'_{2T}$, $\tilde{H}_2$, $H'_2$, $E'_2$, $\tilde{E}'_2$ are $0$. Combining Eqs. (\ref{plusnonconj}) and (\ref{plusconj}) one concludes that, $\tilde{E}_{2T}$, $H'_{2T}$, $E'_{2T}$, $\tilde{H}'_{2T}$, $\tilde{E}_2$, $H_2$, $E_2$, $\tilde{H}'_2$ are real-valued. The same analysis holds for Eqs. (\ref{minusnonconj}) and (\ref{minusconj}). Thus we know that all twist-3 GPDs are real-valued functions.

\section{The OGE potential} \label{oge}
In this Appendix, we present the explicit expression for the effective OGE potential~\cite{Wiecki:2014ola},
% \begin{align}
%     V_{i, j}^{\mathrm{OGE}}=&\sum_n\frac{P^+P^-_{q_i^\prime g\rightarrow q_i}|n\rangle\langle n|P^+P^-_{q_j\rightarrow q_j^\prime g}}{Q^2_{ij}}+P^+P^-_{\mathrm{inst}},
% \end{align}
% \begin{widetext}
    % \begin{align}
    %     V_{i, j}^{\mathrm{OGE}}=&\frac{C_F \alpha_s}{K}
    %     \sum_{s_i,s_i^\prime,s_j,s_j^\prime}\sum_{k_i,k_i^\prime,k_j,k_j^\prime}\delta^{k_i+k_j}_{k_i^\prime+k_j^\prime}
    %     % \sum_{\bar{\alpha}_i,\bar{\alpha}_i^\prime,\bar{\alpha}_j,\bar{\alpha}_j^\prime}\delta^{k^i+k_j}_{k_i^\prime+k_j^\prime}
    %     %\sum_{s_i,s_i^\prime,s_j,s_j^\prime}\int\mathrm{d}x_i\mathrm{d}x_j\mathrm{d}x_i^\prime\mathrm{d}x_j^\prime\delta(x_i+x_j-x_i^\prime-x^\prime_j)
    %     \frac{\int\mathrm{d}^2\vec{p}_{\perp i}\mathrm{d}^2\vec{p}_{\perp j}\mathrm{d}^2\vec{p}^{\;\prime}_{\perp i}\mathrm{d}^2\vec{p}^{\;\prime}_{\perp j}}{(2\pi)^2(2\pi)^2(2\pi)^2(2\pi)^2}\nonumber\\
    %     &\times\frac{(2\pi)^2\delta^2(\vec{p}_{\perp i}+\vec{p}_{\perp j}-\vec{p}_{\perp i}^{\;\prime}-\vec{p}_{\perp j}^{\;\prime})}{(x_i-x_i^\prime)Q_{ij}^2}b_{s_i}^\dagger(p_i)b_{s_j}^\dagger(p_j)b_{s_i^\prime}(p_i^\prime)b_{s_j^\prime}(p_j^\prime)S_{s_i,s_j,s_i^\prime,s_i^\prime}(p_{ i},p_{ j},p_{ i}^\prime,p_{ j}^\prime)
    % \end{align}
    \begin{align}
        &V_{i, j}^{\mathrm{OGE}}=\frac{C_F \alpha_s}{K}\sum_{\alpha_i\alpha_i^\prime\alpha_j\alpha_j^\prime}\delta^{k_i+k_j}_{k_i^\prime+k_j^\prime}b^\dagger_{\alpha_i^\prime}b^\dagger_{\alpha_j^\prime}b_{\alpha_j}b_{\alpha_i}\nonumber\\
        &\times\frac{\int\mathrm{d}^2\vec{p}_{\perp i}\mathrm{d}^2\vec{p}_{\perp j}\mathrm{d}^2\vec{p}^{\;\prime}_{\perp i}\mathrm{d}^2\vec{p}^{\;\prime}_{\perp j}}{(2\pi)^2(2\pi)^2(2\pi)^2(2\pi)^2}\frac{(2\pi)^2\delta^2(\vec{p}_{\perp i}+\vec{p}_{\perp j}-\vec{p}_{\perp i}^{\;\prime}-\vec{p}_{\perp j}^{\;\prime})}{(x_i-x_i^\prime)Q_{ij}^2}\nonumber\\
        &\times \phi_{n_i}^{m_i}(p_{\perp i})\phi_{n_j}^{m_j}(p_{\perp j})\phi_{n_i^\prime}^{m_i^\prime *}(p_{\perp i^\prime})\phi_{n_j^\prime}^{m_j^\prime *}(p_{\perp j}^\prime)\nonumber\\
        &\times (P^+)^2\bar{u}_{\lambda_{i}^{\prime}}\left(p_{i}^{\prime}\right) \gamma^{\mu} u_{\lambda_{i}}(p_{i}) \bar{u}_{\lambda_{j}^{\prime}}\left(p_{j}^{\prime}\right) \gamma_{\mu} u_{\lambda_{j}}\left(p_{j}\right).
    \end{align}
We follow the convention of Ref. \cite{Zhao:2013cma}, where the spinors, $u_\lambda(p)$ and $\bar{u}_\lambda(p)$, are dimensionless. $K$ is the longitudinal truncation. $b^\dagger_\alpha$ and $b_\alpha$ are the creation and annihilation operators in the BLFQ basis~\cite{Zhao:2013cma}. The relations between the normal creation and annihilation operators and the BLFQ ones are given by
\begin{align}
    b^\dagger_\lambda(p)&=\sum_{nm}b^\dagger_\alpha\phi_{n}^{m*}(p_\perp)\\
    b_\lambda(p)&=\sum_{nm}b_\alpha\phi_{n}^m(p_\perp),
\end{align}
where the quantum number set, $\alpha$, includes the longitudinal momentum quantum number $k$, the radial quantum number $n$, the orbital quantum number $m$, and the light-front helicity $\lambda$.

\bibliographystyle{unsrt}
\bibliography{main}

\end{document}